\documentclass[a4paper,11pt]{article}
\pdfoutput=1 

\usepackage{jcappub} 

\usepackage[T1]{fontenc} 
\usepackage{caption}
\usepackage{physics}
\usepackage{mathtools}
\usepackage[lofdepth,lotdepth]{subfig}
\usepackage{subcaption}
\usepackage[normalem]{ulem}

 \newcommand{\be}{\begin{equation}}
 \newcommand{\ee}{\end{equation}}
 \newcommand{\bea}{\begin{eqnarray}}
 \newcommand{\eea}{\end{eqnarray}}

\newcommand{\beq}{\begin{equation}}
\newcommand{\eeq}{\end{equation}}
\let\OldS\S
\renewcommand{\S}{\OldS{}}

\title{Cascades from ultra-high-energy neutrinos}

\author[a,b,1]{Gaetano Di Marco,\note{Corresponding author.}}
\author[c]{Rhorry Gauld,}
\author[d]{Rafael Alves Batista,}
\author[e]{Alfonso García-Soto}
\author[a, b]{and Miguel Á. Sánchez-Conde.}

\affiliation[a]{Instituto de F\'isica Te\'orica UAM/CSIC,\\
Calle Nicol\'as Cabrera 13-15, Cantoblanco, 28049 Madrid, Spain}
\affiliation[b]{Departamento de Física Teórica, M-15, Universidad Autónoma de Madrid, E-28049 Madrid, Spain}
\affiliation[c]{Max-Planck-Institut f\"ur Physik, Boltzmannstraße 8, 85748 Garching, Germany}
\affiliation[d]{Sorbonne Universit{\'e}, Institut d'Astrophysique de Paris (IAP), CNRS UMR 7095\\ 98 bis bd Arago 75014, Paris, France}
\affiliation[e]{Instituto de Física Corpuscular (IFIC), CSIC‐UV, 46980 Paterna, València, Spain}

\emailAdd{gaetano.dimarco@ift.csic.es}
\emailAdd{rgauld@mpp.mpg.de}
\emailAdd{rafael.alves\_batista@iap.fr}
\emailAdd{alfonso.garcia@ific.uv.es}
\emailAdd{miguel.sanchezconde@uam.es}

\def\printinternalnumber{%
    \begin{center}
        \hfill \small\ttfamily MPP-2026-143 
    \end{center}
    \vspace{-5mm} 
}
\makeatother

\date{\today}

\abstract{Neutrinos produced at the highest energies can interact with cosmic neutrino and radiation backgrounds during their propagation to Earth. The many available $\nu\nu$, $\nu\bar{\nu}$, and $\nu\gamma$ channels can lead to their absorption or energy redistribution, whilst the leptonic and hadronic final states may feed secondary fluxes of neutrinos, protons, and electromagnetic particles through the decay or hadronisation of the heavy leptons, bosons, and quarks produced. We present a framework to characterise these propagation effects in detail, \texttt{$\nu$propa}, an extension of the CRPropa Monte Carlo code that interfaces with event generators and to dedicated computations of the relevant cross sections. It also treats flavour oscillations in vacuum. Using this code, we investigate sources at high redshifts ($z = 10$), and find a strong absorption of the prompt flux beyond~$\sim 10^{21} \; \text{eV}$, although the copious secondary neutrinos partially compensate this depletion, also contributing to the spectrum at lower energies. The framework is designed to study scenarios of cosmological neutrino production beyond~EeV energies such as superheavy dark matter, cosmic strings, and primordial black holes, and to yield reliable predictions for the forthcoming neutrino observatories.}

\begin{document}
\printinternalnumber
\maketitle
\flushbottom

\section{Introduction}\label{sec:neutrinoIntro}

Neutrinos are unique messengers of the most extreme phenomena in the Universe, due to their neutral charge and their fleeble interactions with other Standard Model particles. Contrary to cosmic and gamma rays, they can traverse background photon fields and gas without significant absorption, even at extremely-high energies. Neutrino horizons, i.e. the maximum distance a neutrino of a certain energy can reach us unperturbed, are remarkably large over various magnitudes in energy, even the most energetic ones~\cite{berezinsky1992neutrino, roulet1993ultrahigh}. Thus, neutrinos can furnish invaluable insights on the deepest mysteries of the Universe.

Neutrino horizons are evaluated by considering the interaction of highly-energetic neutrinos with low-energy background neutrinos. Among the neutrino pervasive fields~\cite{vitagliano2020grand}, the cosmic neutrino background (C$\nu$B) --- a relic from the early universe freezing-out neutrinos --- is the densest one~\cite{dolgov2002neutrinos, hannestad2006primordial, lesgourgues2013neutrino}. Although it has not been detected so far, its density today is expected to be $\sim 112 \; \text{cm}^{-3}$ per flavour, considering both neutrino and antineutrinos. Their characteristic energies depends on the neutrinos masses, constrained to be $\sum m_{\nu} < 0.45 \; \rm eV/c^2$~\cite{Aker2025KATRIN}, as well as on their characteristic temperature. Furthermore, at energies higher than the \textit{W} boson production energy ($m_{W} \simeq 80 \; \text{GeV}/\text{c}^{2}$), neutrinos might interact with cosmological background photons producing a $W$ boson and lepton final state~\cite{seckel1998neutrino, alikhanov2015glashow, zhou2020neutrino}. As the C$\nu$B for neutrinos, the most relevant relic radiation is the cosmic microwave background (CMB), with a current density of $\sim 411 \; \text{cm}^{-3}$~\cite{aghanim2020planck} and peaked at $\epsilon_{\rm CMB} \sim 6.6 \times 10^{-4} \; \text{eV}$. Because the photons in these cosmological backgrounds are extremely low in energy, only ultra-high-energy neutrinos propagating over cosmological distances are expected to experience significant attenuation through interactions with these relic particles. Moreover, interactions with backgrounds lead to the generation of secondary neutrinos and other particles, as protons, electrons and gamma rays. 

Fluxes of extremely energetic neutrinos, above roughly the EeVs, are expected to be produced by core-collapse supernovae~\cite{janka2012explosion, mirizzi2016supernova, janka2017neutrino} and the interaction of ultra-high-energy cosmic rays with the radiation of the CMB and the extragalactic background light (EBL)~\cite{berezinsky1970origin, allard2006cosmogenic, heinze2016cosmogenic, batista2019cosmogenic}. The latter are the so-called \textit{cosmogenic} neutrinos, considered as the highest energy expected neutrino fluxes within ``standard'' physics frameworks. Alternative scenarios beyond the Standard Model can also give rise to fluxes in this energy domain. Thus, neutrinos roughly exceeding the \textit{cosmogenic} energies would be clear signals of exotic phenomena, such as fluxes generated by topological defects~\cite{berezinsky1998signatures}, Z-bursts~\cite{semikoz2004ultra} or thermally-produced superheavy particles~\cite{kuzmin1999matter, berezinsky2008supersymmetric}, also as candidates for constituting dark matter. 

Such extremely energetic neutrino fluxes constitute the primary science goal of the next-generation detection techniques. Lunar orbital radio observatories as NuMoon~\cite{buitink2008numoonexperimentresults} leverage the Askaryan effect~\cite{dagkesamanskii1989radio} and can detect extremely-energetic neutrinos --- and cosmic rays -- by using radio telescopes~\cite{scholten2009first}. The NuMoon experiment set upper limits on the neutrino fluxes above $\sim 10^{22} \; \text{eV}$, using the LOFAR detectors~\cite{krampah2023numoon}. The FORTE satellite~\cite{lehtinen2004forte} records radio bursts from the Earth surface to detect cosmic-ray particles in the same energy range as NuMoon. Similar planned experiments, e.g. LOFAR~\cite{singh2012optimized} and LUNASKA~\cite{james2010lunaska, james2009sensitivity}, as well as observations with the SKA~\cite{james2017overview} radio telescopes, are expected to reach better detection performances in the near future~\cite{chen2023detection}. In this context, the lunar Ultra-Long Wavelength released the projected all-flavour sensitivities at energies higher than $\sim 10^{21} \; \text{eV}$~\cite{chen2023detection}. Also the~RNO-G project is continuing the deployment of radio antennas in Greenland~\citep{agarwal2025instrument}. In the energy domain from~TeV to several~EeV, many studies on neutrinos production mechanisms have been conducted through the most recent neutrino detections performed by IceCube, that have detected neutrinos up to several~PeV and put upper limits at higher energies~\citep{aartsen2018differential}, and KM3Net, detecting the most energetic neutrino ever to date~\citep{km3net2025observation}. Also the Pierre Auger Observatory has set upper limits beyond hundreds of PeV~\citep{aab2015improved}. Finally, the IceCube-Gen2~\cite{aartsen2021icecube} and KM3Net experiments, covering a broad range from the~GeV to the~EeV domains, and GRAND~\cite{alvarez2020giant}, above $\sim 10^{17} \; \text{eV}$, will significantly improve the current coverage and understanding of the highest energy neutrino sky. With different detection techniques, a similar energy coverage will be possible thanks to the~TRIDENT~\citep{ye2023multi} and~Trinity~\citep{brown2021trinity, bagheri2025commissioning} planned experiments.

This paper aims to show the impact of the propagation effects on the highest-energy neutrino fluxes in a fully-consistent and robust computational framework. As a matter of fact, the full treatment of the (hadronic) cascades arising from the interaction of the energetic neutrinos with the cosmic backgrounds is performed for the first time. It improves the previous analytical studies of propagation effects on neutrinos~\citep{weiler1982resonant, yoshida1994propagation, yoshida1998extremely, barenboim2005diagnostic, das2024probing, wang2026generation} and the calculations neutrino horizons~\citep{eberle2004relic, lunardini2012cosmic, ruffini2016cosmic}. In Sec.~\ref{sec:prop} we describe the theoretical propagation of such energetic neutrinos, defining the sources, backgrounds with their dependence on redshift and mass, elastic and inelastic interactions with photons and other neutrinos and, eventually, the interaction rates as mean free paths. The simulation framework used for the propagation of neutrinos over cosmological distances, the \texttt{$\nu$propa} code\footnote{
  It is available at the \hyperlink{https://github.com/GDMarco/NuPropa/tree/main}{\texttt{$\nu$propa} github page}.
  \raisebox{-0.05\height}{
  \hspace{0.0em}
    \includegraphics[height=1.4em]{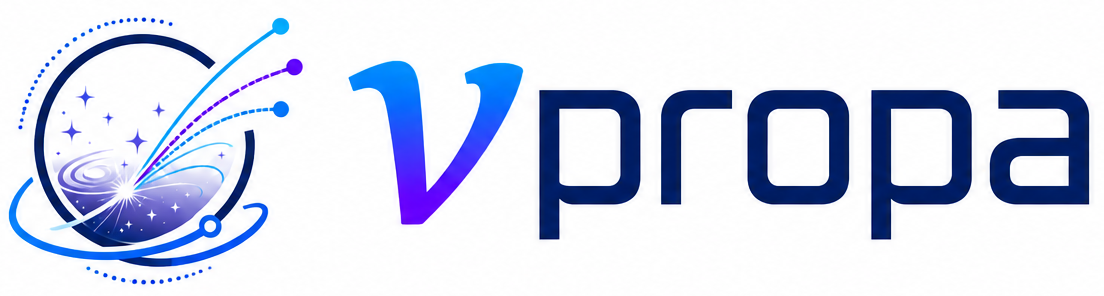}
  }
}, is presented in Sec.~\ref{sec:compFramework}. It is a new plug-in for the Monte Carlo CRPropa code~\cite{Batista_2016, batista2022crpropa} that allows to trace the development of the cascade initiated by energetic neutrinos moving through particle backgrounds. The secondaries from unstable leptons, bosons or quark final states are computed by dedicated interfaces with the PYTHIA code~\cite{bierlich2022comprehensive}, as in Ref.~\cite{di2025gamma}. Lastly, we illustrate how the inclusion and detailed treatment of the propagation effects changes the prompt spectra of extremely energetic neutrinos fluxes (Sec.~\ref{sec:exSim}). Conclusions and broader applicability of this framework are discussed in Sec.~\ref{sec:conclProsp}.   


\section{Propagation of neutrinos}\label{sec:prop}

In this section, we describe phenomena that regards the generation and propagation of extremely-energetic neutrinos. After briefly introducing their production mechanisms and the relic particles permeating the Universe, the processes involving interactions between neutrinos as well as neutrino-photon are introduced. This will constitute the foundation for computing the typical length-scales travelled by neutrinos before interacting with photon and neutrino backgrounds. Finally, the  details regarding the flavour oscillations are described.    

From now on, Greek alphabet indices refer to the neutrino flavours eigenstates, viz. $\alpha = \{e,\mu , \tau\}$, while Roman ones are used for the mass states, $i = \{1, 2, 3\}$.


\subsection{Production of extremely-energetic neutrinos}\label{sec:nuProd}
Energetic neutrinos, roughly beyond the~GeVs, have been found to be produced by standard astrophysical phenomena and, in some exotic scenarios, are expected from beyond Standard Models theories. In the GeV-to-PeV energy regime, the production of neutrinos is usually attributed to various classes of extreme extragalactic and galactic sources~\cite{murase2015origin}. Among the extragalactic ones, active galactic nuclei, supernovae and gamma-ray bursts are efficient cosmic-ray accelerators, provided by local environments suitable for (photo)hadronic production of neutrinos. Their escaping cosmic rays can still produce neutrinos, especially if these sources are embedded in efficient reservoirs as galaxy clusters~\cite{hussain2021high} and starburst galaxies~\citep{LoebWaxman2006Starburst, tamborra2014star}. In the last decade, just a few of the many GeV-TeV neutrinos detected by the IceCube experiment have been found to be associated to active galactic nuclei~\citep{icecube2018multimessenger, evidence2022}. In our galaxy, as well as in many others that could feed clusters, cosmic rays from local accelerators as supernova remnants, pulsar wind nebulae and microquasars interacts within sources themselves or in the interstellar medium to produce neutrinos. Recently, the IceCube collaboration reported a clear association with the galactic plane, although no significant association to sources has been found~\cite{icecube2023observation}. At even higher energies, the so-called \textit{cosmogenic} neutrino fluxes are predicted by the most energetic cosmic rays, escaping from their sources or surrounding reservoirs, by interacting with the CMB in the extragalactic space~\cite{greisen1966end, Zatsepin:1966jv, 1969PhLB...28..423B, stecker1973ultrahigh}. The models predict neutrinos beyond~EeV energies~\cite{kotera2010cosmogenic, batista2019cosmogenic, alves2019open}, depending on the source distribution and acceleration efficiency. Such fluxes, though, are not accessible to present-day experiments~\citep{kotera2026a}. The production of cosmogenic neutrinos with energies beyond~$\sim 10^{21}-10^{22}\;\mathrm{eV}$ is expected to be highly unlikely~\cite{thompson2011upper, heinze2016cosmogenic, batista2019cosmogenic}.

Energetic neutrinos are also predicted to arise in several exotic scenarios. In fact, GeV--PeV neutrino fluxes could be generated from the annihilation or decay of weakly interacting massive particles, while beyond few~PeVs from super heavy dark matter particles, whose masses can reach the Planck scale. Both dark matter candidates have been constrained by state-of-the-art facilities observing neutrinos~\citep{chianese2021heavy, song2024search, kohri2025super, adriani2026testing} and other messengers~\citep{ishiwata2020probing, auger2022c, das2023revisiting, auger2023c, cao2024constraints}. Projected constraints on the properties of these dark matter candidates have been estimated for future neutrino experiments~\citep{das2024probing} that will reach extremely-high energies, surpassing the $10^{20} \; \text{eV}$. Beyond dark matter, such energetic fluxes of cosmic neutrinos can also be produced from the interaction of scalar fields with primordial cosmic strings~\cite{bhattacharjee1990ultrahigh, berezinsky2011extremely, lunardini2012cosmic, creque2023high}, also superconducting~\cite{witten1985superconducting, berezinsky2009ultrahigh}. Primordial black holes constitute yet another possibility~\cite{PhysRevD.52.3239, Lunardini_2020, wu2025high}. More generally, such energetic neutrinos can be produced by any decaying primordial relic decaying directly in different $\nu$ states~\cite{PhysRevLett.44.1481, GONDOLO1993111, PhysRevD.76.105017, yamamoto2025primordial}. Some of these generic models have already been constrained using data from from Ice~Cube~\cite{EMA2014120} and KM3Net~\cite{yamamoto2025primordial}.

Once produced, these neutrinos fluxes, characterised by a specific prompt spectral shape, according to the particular scenario considered, begin travelling towards the Earth. Throughout their journey, diffuse backgrounds of low-energy neutrinos and photons might constitute interaction targets depending on the propagating neutrino energy and distance travelled. These interactions can modify the initial prompt energy spectra, leaving characteristic imprints on the observed fluxes. Depending on the specific production model, the potential derivation of their flavour composition, combined with spectral features, could be strikingly indicative of the underlying generation model convolved with the propagation effect~\citep{coleman2024flavor}.


\subsection{Low-energy cosmic backgrounds}\label{sec:cosmBackg}
\begin{figure}
    \centering

    \begin{minipage}{0.48\linewidth}
        \centering
        \includegraphics[width=\linewidth]{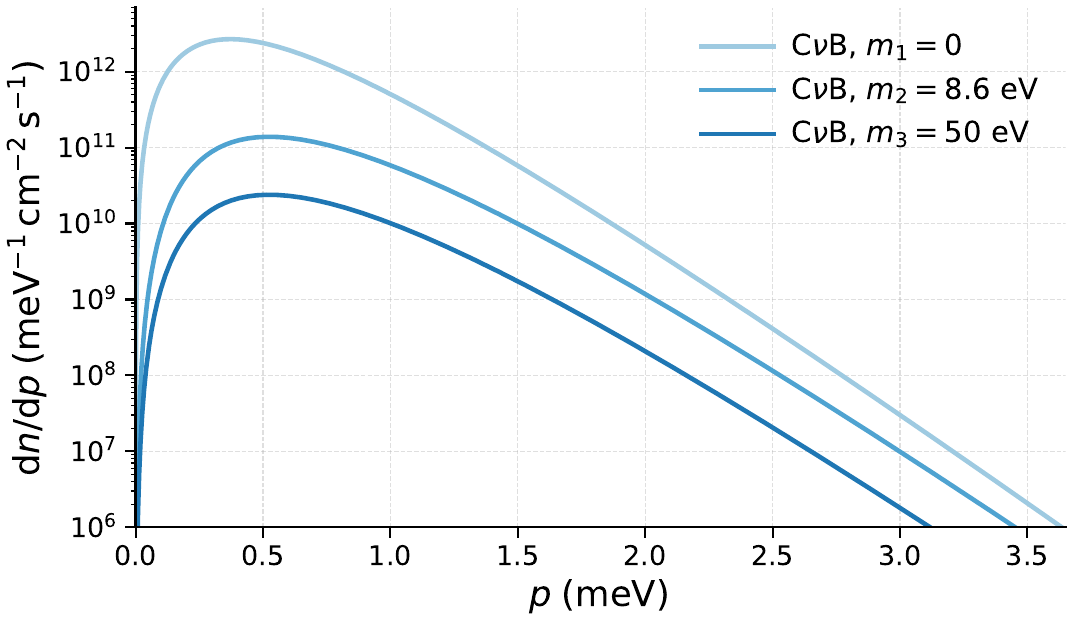}
        
        \vspace{0.2cm}
        {\small C$\nu$B fluxes in terms of the particle momenta~$p$, as given by Eq.~\eqref{eq:bkgMomDistr}.}
        \label{fig:densityMom}
    \end{minipage}
    \hfill
    \begin{minipage}{0.48\linewidth}
        \centering
        \includegraphics[width=\linewidth]{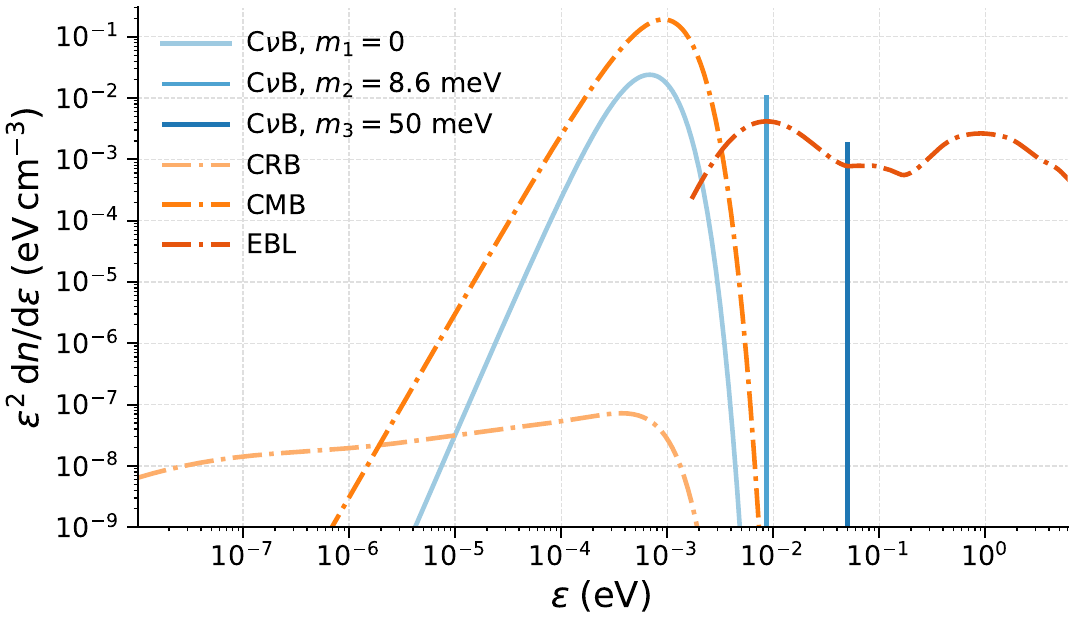}
        
        \vspace{0.2cm}
        {\small Energy spectra of the cosmic backgrounds. The~CRB and~EBL are from Refs.~\cite{nictu2021updated} and~\cite{saldana2021observational}.}
        \label{fig:backgroundPhotonNeutrino}
    \end{minipage}

    \caption{
    Densities of the cosmic backgrounds at redshift $z=0$ in terms of the momenta (left) and of the energies (right). The latter represents both the neutrino and photon backgrounds, respectively the solid and dashed-dotted lines.
    }
    \label{fig:CosmicBackground}
\end{figure}

$\Lambda$CDM standard cosmology predicts the existence of the C$\nu$B and the CMB originating from the epochs of neutrino decoupling and photon decoupling, respectively. It is worth emphasizing that thermal effects of the C$\nu$B are relevant only when the background neutrinos are non-relativistic, i.e. if the neutrino mass is comparable or smaller than the temperature of the background black-body spectrum. This condition is satisfied at $z \lesssim 83$, taking $m \gtrsim 0.05 \; \text{eV}$ as the heaviest neutrino mass~\cite{lunardini2013ultra}.
Thus, the C$\nu$B energy spectrum depends on the neutrino energy $\epsilon$ and the redshift $z$. Assuming an isotropic background of neutrinos with mass $m_{\mathrm{bkg}}$, the density distribution can be expressed in terms of the momentum~$p$ as~\cite{alpher1953physical}:
\begin{equation}\label{eq:bkgMomDistr}
    \dv{n_{\nu_{i}+\bar{\nu}_{i}}}{p} \left(p, z \right) = \frac{4\pi}{c^{3}h^{3}}\cdot \frac{p^{2}c^{2}}{\text{exp}\left( \frac{pc}{k_{B}T_{0\nu}(1 + z)}\right) + 1} \,,
\end{equation}
with $T_{0\nu}$ the \textit{current} characteristic temperature ($c$ is the speed of light, $h$ and $k_{B}$ the, respectively, Planck and Boltzmann constants). This distribution, properly converted into a flux by accounting for the relativistic particle velocity, results in the density profiles of Fig.~\ref{fig:CosmicBackground}~(left). 

Other relic low-energy neutrinos are expected from Big Bang nucleosynthesis as products of neutron and tritium decays. However their densities are much lower than the C$\nu$B ones~\cite{khatri2011time, ivanchik2018relic, vitagliano2020grand}.

The isotropic CMB density spectrum, with \textit{current} characteristic temperature $T_{0\gamma}$, is described by the Bose-Einstein distribution~\citep{bose1924plancks, einstein2005quantentheorie}:  
\begin{equation} \label{eq:bkgEDistr}
    \dv{n_{\gamma}}{\epsilon}\left(\epsilon, z\right ) = \frac{8\pi}{c^{3}h^{3}}\cdot \frac{\epsilon^{2}}{\text{exp}\left( \frac{\epsilon^{2}}{k_{B}T_{0\gamma}(1 + z)}\right ) - 1} \,.
\end{equation}

Apart from the relic CMB, other photons have been permeating the intergalactic space over various cosmic times (a general review can be found in Ref.~\cite{cooray2016extragalactic}). The lowest energy one is the cosmic radio background (CRB) and it is the result of radio emission from astrophysical objects, following the history of formation and evolution~\cite{protheroe1996new, nictu2021updated}. The equivalent photon field between the infrared and the ultraviolet is the~EBL~\cite{franceschini2008extragalactic, finke2010modeling,dominguez2011extragalactic, gilmore2012semi, stecker2016empirical, saldana2021observational, finke2022modeling}. The energy spectra of the these neutrino and photon backgrounds are reported in Fig.~\ref{fig:CosmicBackground} (right). Since in this work we focus on the extremely energetic neutrinos, we will only consider the lowest energy neutrino and photon backgrounds (which are also the densest). 

\subsection{Interactions involving neutrinos}\label{sec:Interactions}

\begin{figure}
    \centering
    \includegraphics[width=0.75\linewidth]{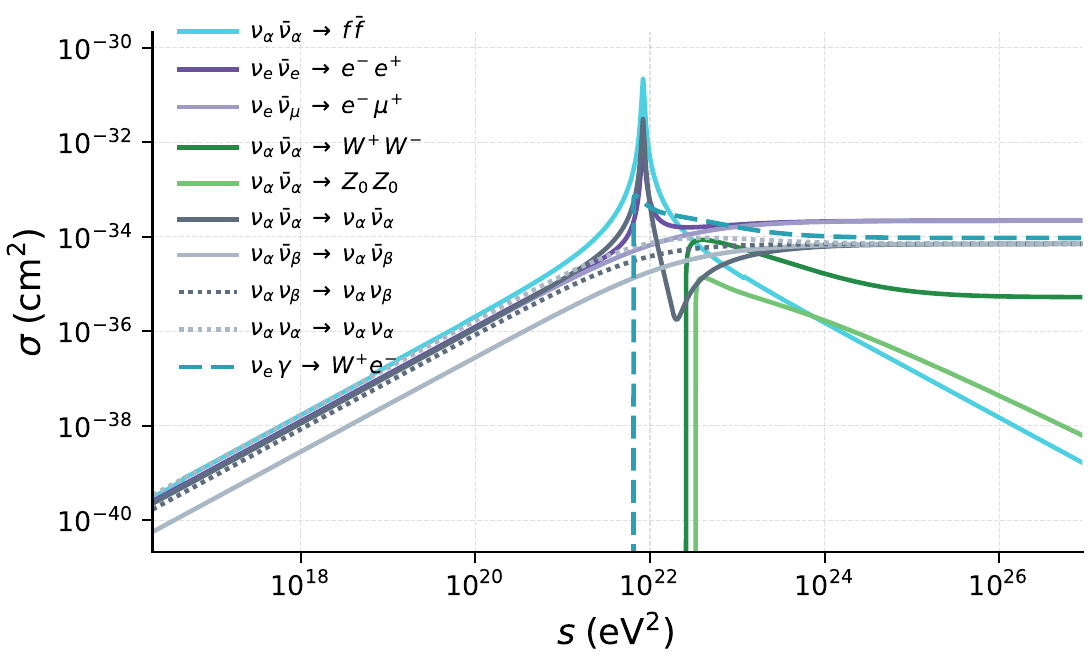}
    \caption{Cross section as function of the centre-of-mass energy for interactions between (anti)neutrinos of different flavours~\cite{berezinsky1992neutrino}, as well as the on-shell W-boson production in neutrino-photon interactions.}
    \label{fig:CS_neutrino}
\end{figure}

Assuming a standard \(\Lambda \)CDM cosmology, space is permeated by low-energy neutrino and photon backgrounds, as described by Eqs.~\eqref{eq:bkgMomDistr} and \eqref{eq:bkgEDistr}. 
When high-energy neutrinos traverse cosmic distances, they may scatter off these backgrounds through various channels~\cite{weiler1982resonant,weiler1984big,roulet1993ultrahigh, yoshida1994propagation,weiler1999cosmic,fargion1997ultrahigh}. These interactions can modify the primary neutrino's energy and trajectory via elastic scattering, induce annihilation, or generate secondary neutrinos.
Given the Standard Model Lagrangian, one can exhaustively enumerate all relevant interaction channels at fixed-order in electromagnetic coupling $\alpha$.
Working at leading-order (LO), while making some assumptions about the treatment of massive gauge bosons, the main interaction channels can be easily identified by considering their total cross-section rates as a function of the squared centre-of-mass energy~$s$.
%
The cross sections of the various processes considered in this work (see also~\cite{seckel1998neutrino}) are summarised in Fig.~\ref{fig:CS_neutrino} in a region of relatively large $s$.

These processes involve "neutrino-photon" interactions between high-energy neutrinos and background photons, as well as "neutrino-neutrino" or "neutrino-antineutrino" interactions with background (anti)neutrinos.
%
Each of these interaction types can be further categorised by neutrino flavour.

Further details of the cross section model that has been adopted in this work, together with details of its implementation for the description of differential scattering processes, are presented in Section~\ref{sec:dsigma}.
In the following, we briefly discuss the main features of the different neutrino-interaction channels.

\paragraph{Neutrino-neutrino ($\nu_{\alpha}\nu_{\beta}$) interactions.}
The interactions between neutrinos or antineutrinos is described by an elastic scattering process.
Inspection of Fig.~\ref{fig:CS_neutrino} indicates that, in the region of low $s$, the same flavour channel has a relatively large cross-section as compared to the non-equal flavour channels.
Instead, in the limit of large $s$, the various channels tend to the same value.
When occurring in the cosmological propagation of neutrinos, these scattering processes act to redistribute the neutrino energies while conserving the total energy of the neutrino population. 

\paragraph{Neutrino-antineutrino ($\nu_{\alpha}\bar{\nu}_{\beta}$) interactions.} 
As in the case of neutrino-neutrino interactions discussed above, there is a similar elastic scattering channel for the neutrino-antineutrino of the form $\nu_{\alpha}\bar{\nu}_\beta\to \nu_{\alpha}\bar{\nu}_\beta$.
Other interesting, and numerically larger, interaction channels are those which involve the annihilation of either a flavour correlated $(\alpha=\beta)$ or uncorrelated ($\alpha\neq\beta$) neutrino-antineutrino pair.

As shown in Fig.~\ref{fig:CS_neutrino}, the largest contribution the cross-section arises from the channel $\nu_{\alpha} \bar \nu_{\alpha}\to f \bar f$, which receives a resonant enhancement (i.e. through an $s$-channel $Z^{0}$-exchange) in the vicinity of $s \sim m_Z^2$. A special case exists when the outgoing fermion pair is composed of a neutrino-antineutrino, i.e. $\nu_{\alpha} \bar \nu_{\alpha} \to \nu_{\alpha} \bar \nu_{\alpha}$ which also receives a $t$-channel contribution.

Other relevant channels have the form $\nu_{\alpha}\bar{\nu}_{\beta} \to \ell_{\alpha} \bar \ell_{\beta}$ and are mediated through a $t$-channel $W^{\pm}$ boson exchange.
Again, a special case exists when $\alpha = \beta$, that also involves a resonant enhancement for $s \sim m_Z^2$. These channels have a large cross-section in the region of $s \gtrsim m_Z^2$. Above the kinematic threshold of $s \gtrsim (2\,M_{V})^{2}$, $V=W,Z$, it becomes kinematically possible to produce a massive gauge boson pair.
The impact of these contributions is shown as light and dark green lines in Fig.~\ref{fig:CS_neutrino}. Through decay, these channels may lead to a source of secondary neutrinos. 

\paragraph{Neutrino-photon ($\nu_{\alpha}\gamma$) interactions.}

The scattering rate of a neutrino and a photon is typically negligibly small as compared to the various neutrino-neutrino and neutrino-antineutrino channels discussed above.

An exception occurs at high-energy ($s \gtrsim m_W^2$) where it becomes kinematically possible to resonantly produce a $W$ boson through the trident scattering process.
In general, the trident production process with an on-shell photon is a $2\to3$ process~\cite{gaidaenko2001production, vysotsky2002lepton, altmannshofer2014neutrino}.
However, as the cross section is numerically dominated by the resonant contribution, it becomes possible to approximate the trident scattering process as $\nu + \gamma \to \ell + W$~\cite{seckel1998neutrino, alikhanov2015glashow, zhou2020neutrino}--- see also the discussion in Appendix~\ref{app:tridentNuPhoton}.

Inspection of Fig.~\ref{fig:CS_neutrino} indicates that at energies close to the $Z$ boson resonance and below, the cross section resulting from neutrino-neutrino interactions is dominant. However, at higher energies the neutrino-photon interaction is comparable or even larger than those from neutrino-neutrino interactions. 
  
\subsection{Mean free paths of cosmological neutrinos}

Given the extremely energetic neutrino sources, the interactions in play and the relic backgrounds, the interactions inverse mean free paths, denoted as $\lambda^{-1}$, can be readily computed. For the interaction of a $\alpha$-flavoured neutrino of energy $E$ with a neutrino background of mass $m_{\text{b}}$ is convenient a rate calculation in the background momentum space~(the distribution in Eq.~\eqref{eq:bkgMomDistr}), thus:   

\begin{equation}\label{eq:IMFPmom}
\lambda_{\alpha\beta}^{-1} \left(E,z ; m_{\text{b}}, m_{\nu}\right )=\frac{1}{8E^{2}} \int\limits_{p_{\text{min}\left (E \right )}}^{\infty}\dd p \, \frac{1}{\beta_{\nu}cp \sqrt{p^{2}c^{2}+m_{\text{b}}^{2}c^{4}}} \dv{n}{p} \left(p,z \right) \int\limits_{s_{\text{min}}}^{s_{\text{max}}}\dd s \, \beta_{\text{rel}}(s) \mathcal{F}_{\alpha\beta}(s) \,.
\end{equation}

Note that the flavour of the background neutrinos, $\beta$, and the mass of the propagating neutrino, $m_{\nu}$, are taken from the flavour-mass matrix probability~(introduced in next section, Sec.~\ref{sec:oscillation}) in order to set, respectively, the interaction allowed cross sections and kinematic factors. The validity of this assumption will become clear once the spirit of the code implemented in this work is explained in~Sec.~\ref{sec:nupropa}. Setting $\beta_{\nu} = 1$, as is the case for the very energetic propagating neutrinos considered here, the minimum allowed momentum for the integration is:
\begin{equation}
    cp_{\text{min}} \left (E_{\nu} \right ) = \frac{K(E_{\nu})^{2}-m_{\text{b}}^{2}c^{4}}{2 K(E_{\nu})} \,,
\end{equation}
with the parameter dependent on the propagating neutrino energy defined as:
\begin{equation}
    K\left (E_{\nu} \right ) = \frac{s_{\text{thr}} - (m_{\nu}^{2} + m_{\text{b}}^{2})c^{4}}{2E_{\nu}} \,. 
\end{equation}
As usual, the interaction threshold is defined by the $N$ interaction products' masses, hence $s_\text{thr} = \sum_{i=0}^{N}m_{i}^{2} c^{4}$. This latter corresponds to the $s$ integration minimum of Eq.~\eqref{eq:IMFPmom}. The relative velocity between the two interacting particles, resulting from their 4-momenta, can be written in terms of the squared centre-of-mass energy: 
\begin{equation}
    \beta_{\text{rel}}(s)= \sqrt{1-4\bigg(\frac{m_{\nu}m_{\text{b}}c^{4}}{s -(m_{\nu}^{2}+m_{\text{b}}^{2})c^{4}}\bigg)^2} \,.
\end{equation}
Generally, for large $s$ and small neutrino masses, the relative velocity is usually close to unity. Lastly, the $\mathcal{F}(s)$ function encodes the process cross section, according to the neutrinos' flavours:
\begin{equation}
   \mathcal{F}_{\alpha\beta} = \left(s - \left(m_{\text{b}}^{2}+m_{\nu}^{2}\right) c^{4}\right) \sigma_{\alpha\beta}(s) \,.
\end{equation}
If both neutrinos lies in a massless ground state, the interaction rates of Eq.~\eqref{eq:IMFPmom} collapse to the one used to compute photon-photon interaction (see, for instance, Ref.~\cite{alves2021gamma}). The same computation is used to derive the inverse mean free paths for neutrino-photon processes.

\begin{figure*}[t]
    \centering
    
    \subfloat[For neutrinos travelling through the~CMB and~EBL (shadowed).\label{fig:IMFPnuGamma}]
    {\includegraphics[width=0.32\textwidth]{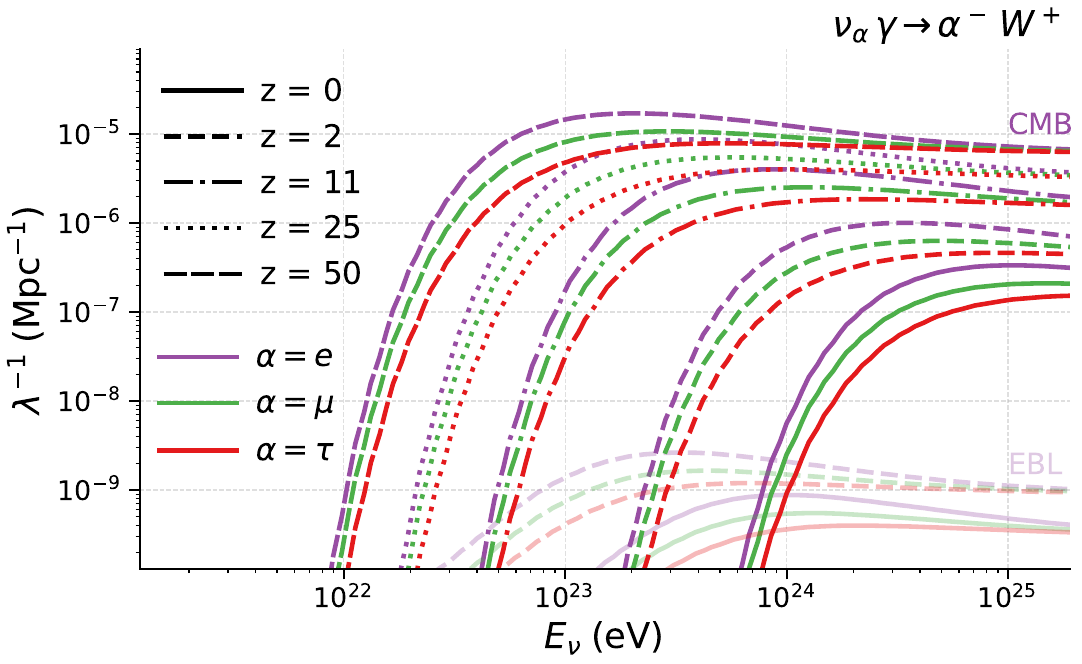}}
    \hfill
    \subfloat[For the $u\bar{u}$ interaction channel.\label{fig:IMFPnunubarTouubar}]
    {\includegraphics[width=0.32\textwidth]{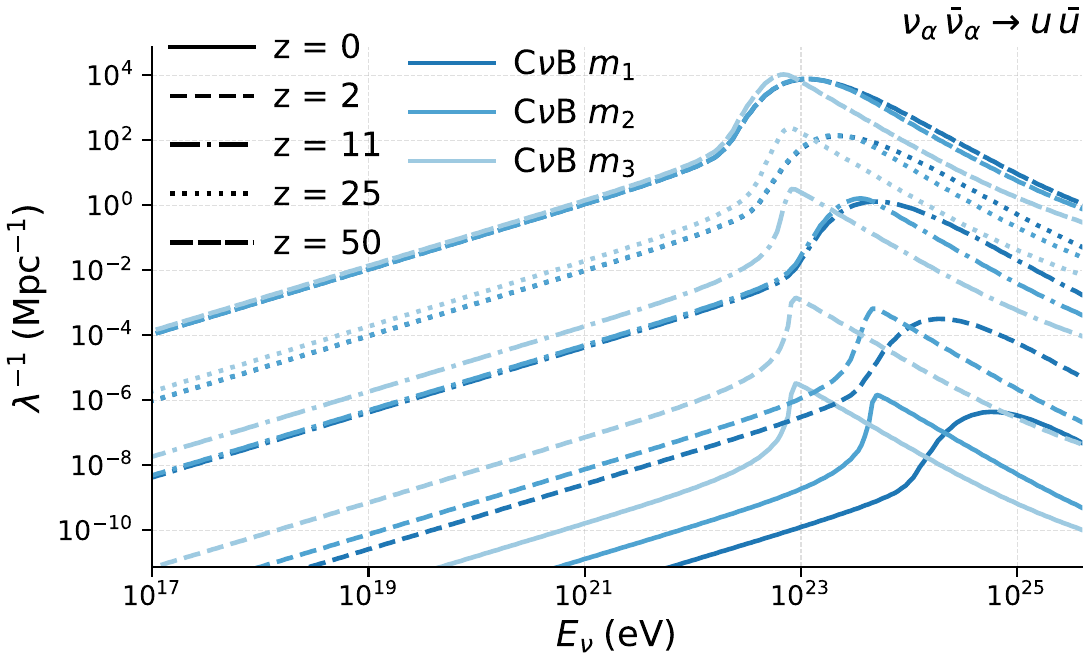}}
    \hfill
    \subfloat[For the elastic $\nu\nu$ interaction channel.\label{fig:IMFPnunuTonunu}]
    {\includegraphics[width=0.32\textwidth]{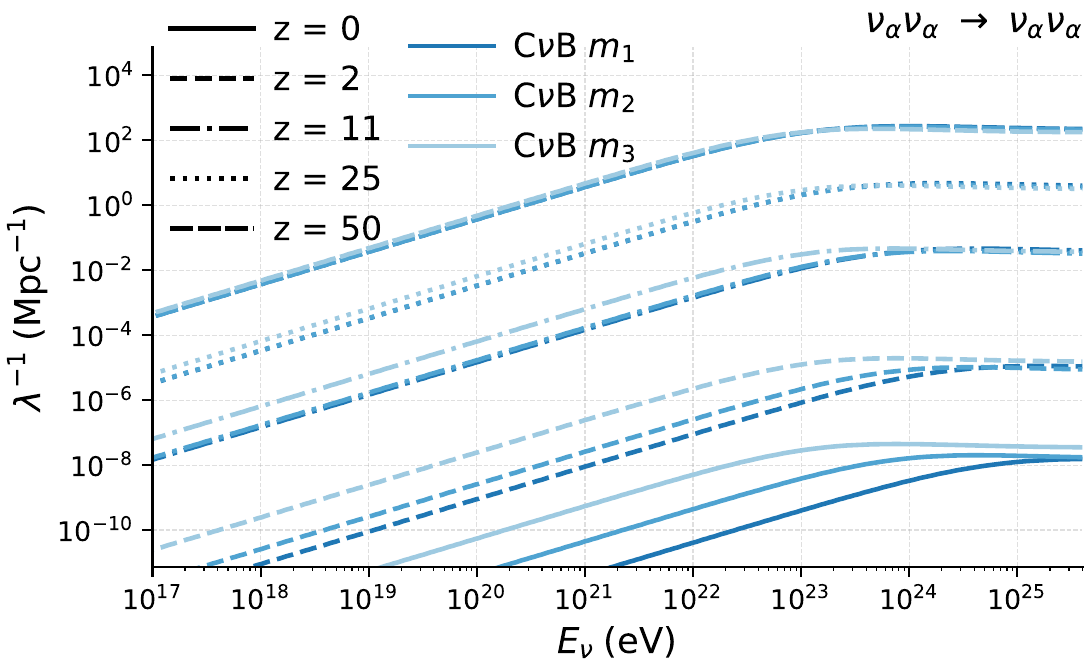}}

    \medskip

    \subfloat[For the $W$ pair production channel.\label{fig:IMFPnunubarToWW}]
    {\includegraphics[width=0.32\textwidth]{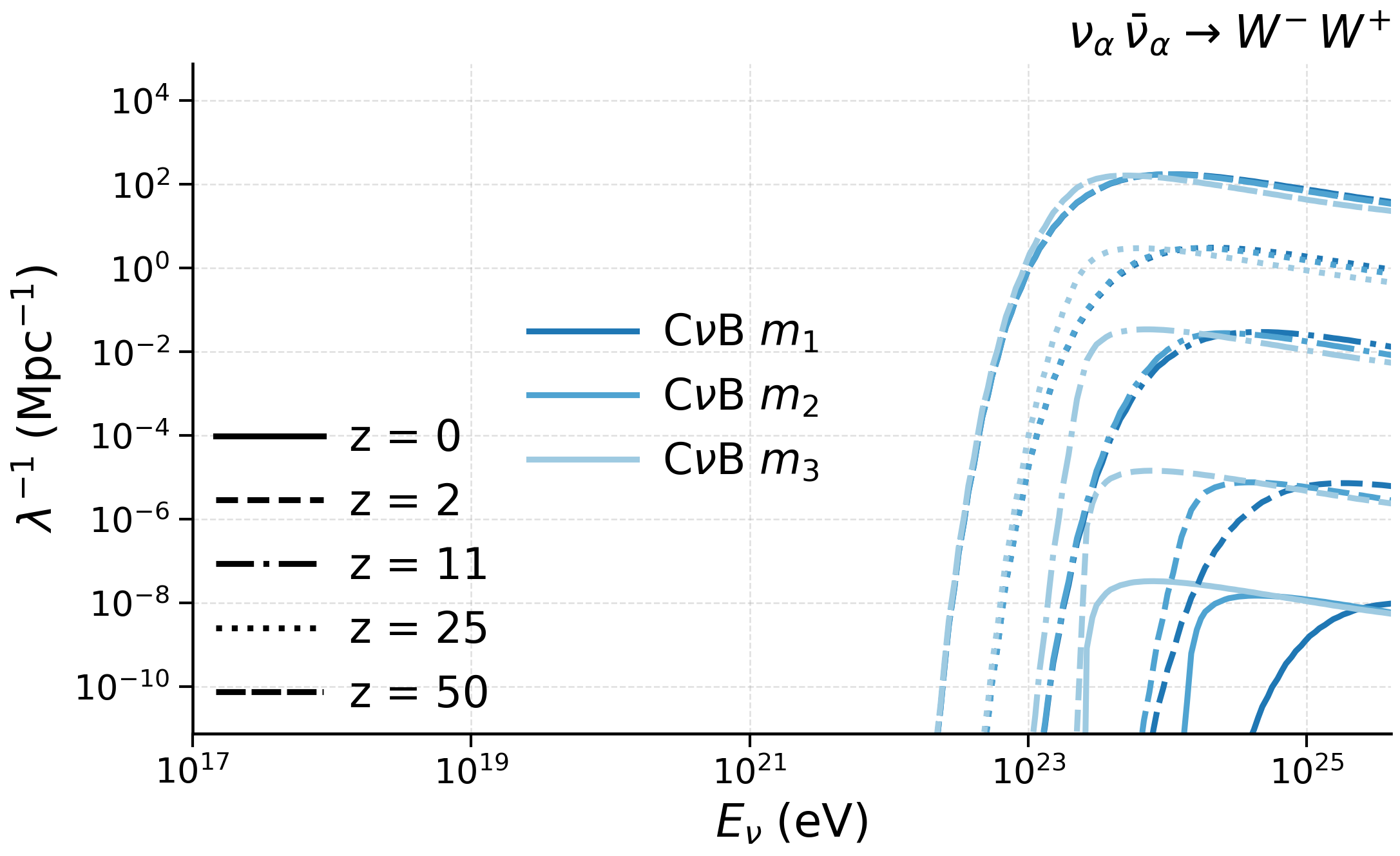}}
    \hfill
    \subfloat[For the elastic interaction of different flavour $\nu\bar{\nu}$.\label{fig:IMFPnuinujbarTonuinujbar}]
    {\includegraphics[width=0.32\textwidth]{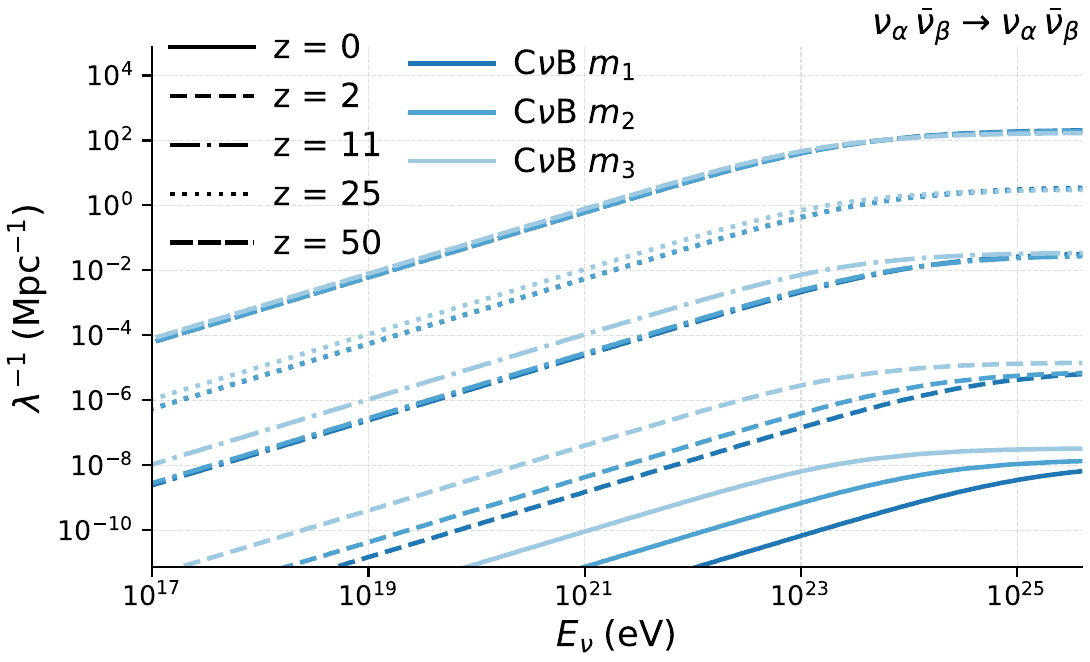}}
    \hfill
    \subfloat[For a specific leptonic production channel.\label{fig:IMFPnuinujbarToElAmu}]
    {\includegraphics[width=0.32\textwidth]{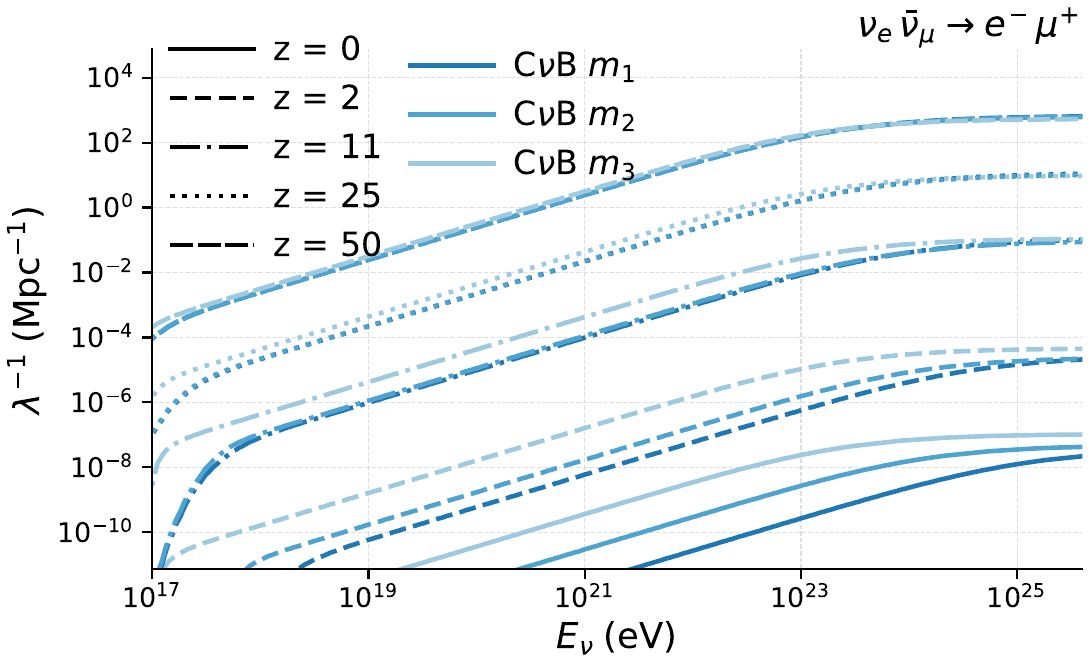}}

    \caption{The inverse mean free paths of some interaction channels that involve a neutrino of energy $E_{\nu}$ propagating through the cosmic either neutrino or microwave background (see Sec.~\ref{sec:cosmBackg}).}
    \label{fig:IMFP_all}
\end{figure*}

The resulting interaction rates for extremely-energetic neutrinos interacting with~CMB and~EBL photons are shown in Fig.~\ref{fig:IMFPnuGamma}. The rates for interacting with the background photons to the different neutrino flavours and redshifts, up to 50, are displayed. The rates for the~EBL are presented up to redshift~2 because of the large uncertainties on their densities at later cosmological epochs. Only the highest-energy peak of the~EBL (see the plot on the right of Fig.~\ref{fig:CosmicBackground}) emerges in the rates of Fig.~\ref{fig:IMFPnuGamma}. The infrared radiation of the~EBL would affect the propagation of lower energy neutrinos, although being much lower. Fig.~\ref{fig:IMFPnuGamma} indicates the irrelevance of the~EBL in the propagation of neutrinos. Different channel rates with the three mass/flavour states of the backgrounds are shown in Figs.~\ref{fig:IMFPnunuTonunu},~\ref{fig:IMFPnunubarToWW},~\ref{fig:IMFPnuinujbarTonuinujbar} and~\ref{fig:IMFPnuinujbarToElAmu} for the same redshifts as before. To evaluate the effectiveness of these processes such rates have to be compared to the inverse of the Hubble radius. This latter is defined at present time as $R_{0}\equiv H_{0} c^{-1}\sim 4400 \; \text{Mpc}$, considering $H_{0} = 67.77 \; \text{km} \; \text{Mpc}^{-1} \; \text{s}^{-1}$ the Hubble constant and $c$ the speed of light. 

\subsection{Neutrino mass--flavour mixing $\&$ oscillation}\label{sec:oscillation}

The probabilities used to draw the neutrino masses from flavour states and viceversa are computed through the Pontecorvo-Maki-Nakagawa-Sakata~(PMNS) matrix~\cite{AMSLER20081}:
\begin{equation}\label{eq:PMNSmatrix}
U_{\text{PMNS}} =
\begin{pmatrix}
c_{12} c_{13} &
s_{12} c_{13} &
s_{13} e^{-i\delta} \\[6pt]
- s_{12} c_{23} - c_{12} s_{23} s_{13} e^{i\delta} &
\;\; c_{12} c_{23} - s_{12} s_{23} s_{13} e^{i\delta} &
s_{23} c_{13} \\[6pt]
\;\; s_{12} s_{23} - c_{12} c_{23} s_{13} e^{i\delta} &
- c_{12} s_{23} - s_{12} c_{23} s_{13} e^{i\delta} &
c_{23} c_{13}
\end{pmatrix}
 \,,
\end{equation}
with $c_{ij} = \cos\theta_{ij}$ and $s_{ij} = \sin\theta_{ij}$. The oscillation parameters, thus $\theta_{ij}$ and the CP-violating phase $\delta$, whose most recent derived values can be found in Ref.~\cite{esteban2024nufit}. Through the~PMNS matrix a $\alpha$-flavoured neutrino can be expressed as superposition of mass states:
\begin{equation}\label{eq:flavMassEig}
 \ket{\nu_{\alpha}} = \sum_{k=1}^{3} U_{\alpha k}^{*}\ket{\nu_{j}} \,.
\end{equation}
In a similar way, neutrino mass states can be expressed as a linear combination of the three flavour ones. By bracketing such flavour states with the mass ones, the transition probabilities are readily obtained. 

Furthermore, since neutrinos are massive --- at least two out of three --- and lepton flavour is not conserved in charge-current interaction, the neutrino flavour states of Eq.~\eqref{eq:flavMassEig} evolve while propagating through the extragalactic \textit{quasi}-vacuum space. The flavour state evolution is expressed as a combination of time-dependent mass states that, in the plane wave assumption, take the form: $\ket{\nu_{i}(t)}=e^{-iE_{i}t}\ket{\nu_{i}(0)}$. Thus, it depends on the neutrino energy and distance covered, respectively $E$ and $L\backsimeq ct$, and, lately, on the mass splitting, having $\braket{\nu_{i}}{\nu_{j}}=\delta_{ij}$. The exact flavour oscillation formula~\cite{PDGNeutrino2022}: 

\begin{equation}\label{eq:oscilProb}
   P_{\nu_\alpha \to \nu_\beta} =
\delta_{\alpha\beta}
- 4 \sum_{i>j}
\Re\!\left(
U_{\alpha i} U_{\beta i}^*
U_{\alpha j}^* U_{\beta j}
\right)
\sin^2\!\left(
\frac{\Delta m_{ij}^2 L}{4E}
\right)
+ 2 \sum_{i>j}
\Im\!\left(
U_{\alpha i} U_{\beta i}^*
U_{\alpha j}^* U_{\beta j}
\right)
\sin\!\left(
\frac{\Delta m_{ij}^2 L}{2E}
\right) \,. 
\end{equation}

Note that, for large arguments of the sines functions, the probability above starts to rapidly oscillating. The behaviours of the oscillation probabilities are exhibited in Fig.~\ref{fig:oscPlot} in dependence to the neutrino energy for a initial muonic state and a~$1 \; \text{pc}$ baseline. The transition to the \textit{fast oscillating} regimes and the average probabilities are highlighted.  

Although not treated in this work, it is worth to mention an additional effect that might derive, under certain circumstances depending on the assumptions, from the propagation over long distances of neutrinos. Since their flavour states are described by a superposition of mass eigenstates (see Eq.~\eqref{eq:flavMassEig}), which may propagate with different velocities, leading to \textit{decoherence}. This phenomena is expected to generate additional phases in the oscillation pattern of Eq.~\eqref{eq:oscilProb} and additional propagation effects~\cite{eberle2004relic, farzan2008coherence}.

\section{Simulations of cosmological neutrinos}\label{sec:compFramework}
To study the scenario in which energetic neutrinos propagates through particle backgrounds, we implement a dedicated plug-in for the CRPropa~3.2 code~\cite{Batista_2016, batista2022crpropa}. 
In the CRPropa framework, Monte Carlo simulations are performed to propagate highly energetic particles over cosmological distances, including all the relevant backgrounds and interactions, as well as the emission properties and the observer characteristics.

\subsection{The \texttt{$\nu$propa} plug-in}\label{sec:nupropa}

\begin{figure}
    \centering
    \includegraphics[width=0.95\textwidth]{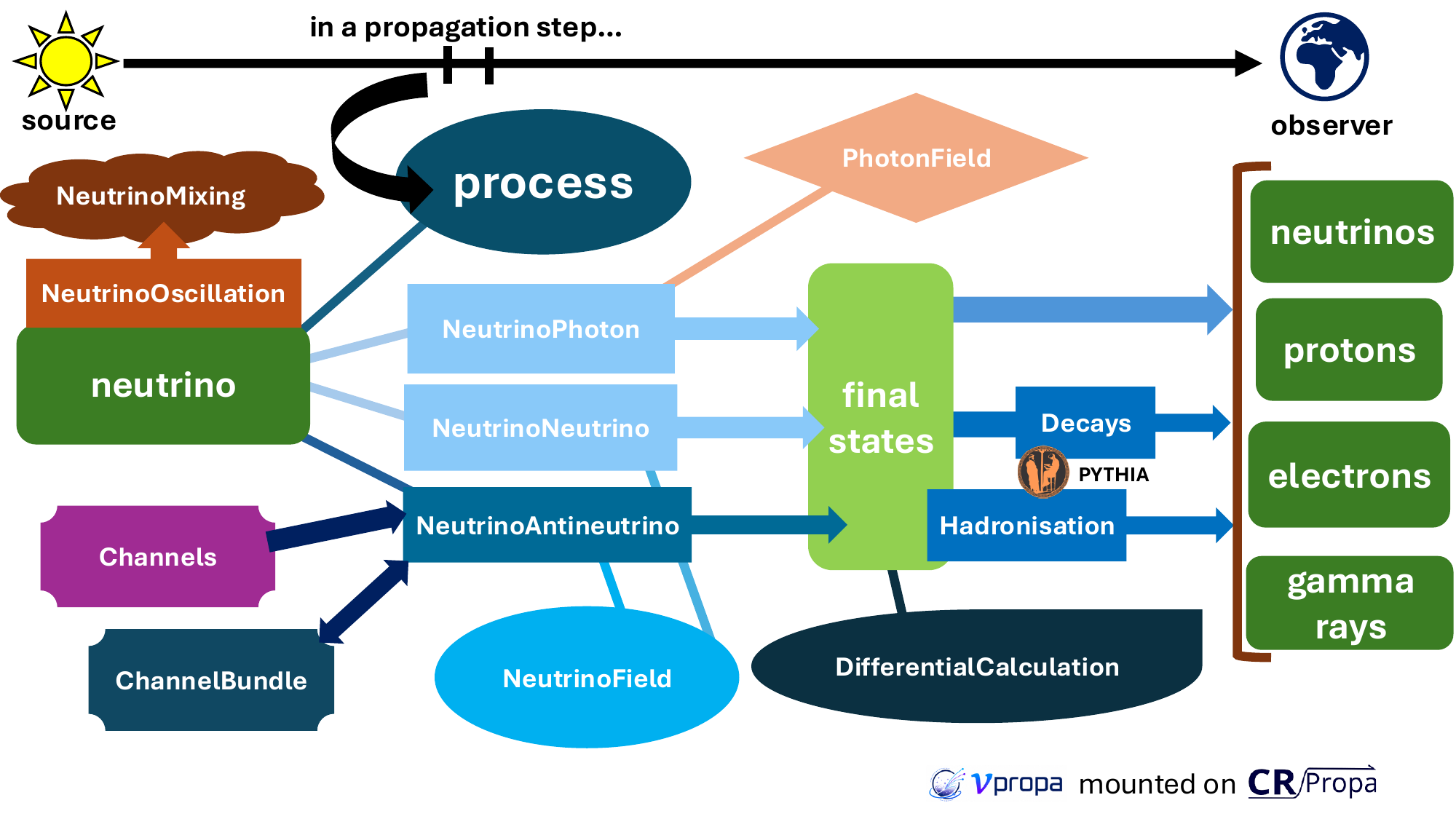}
    \caption{Chart on the workflow of the~\texttt{$\nu$propa} code. The blocks represent the various modules, classes or frameworks of the code. The \texttt{process} function is the main one in~CRPropa: it acts on the propagating particle according to the active modules.}
    \label{fig:chart}
\end{figure}

\begin{figure}
    \centering
    \includegraphics[width=0.46\textwidth]{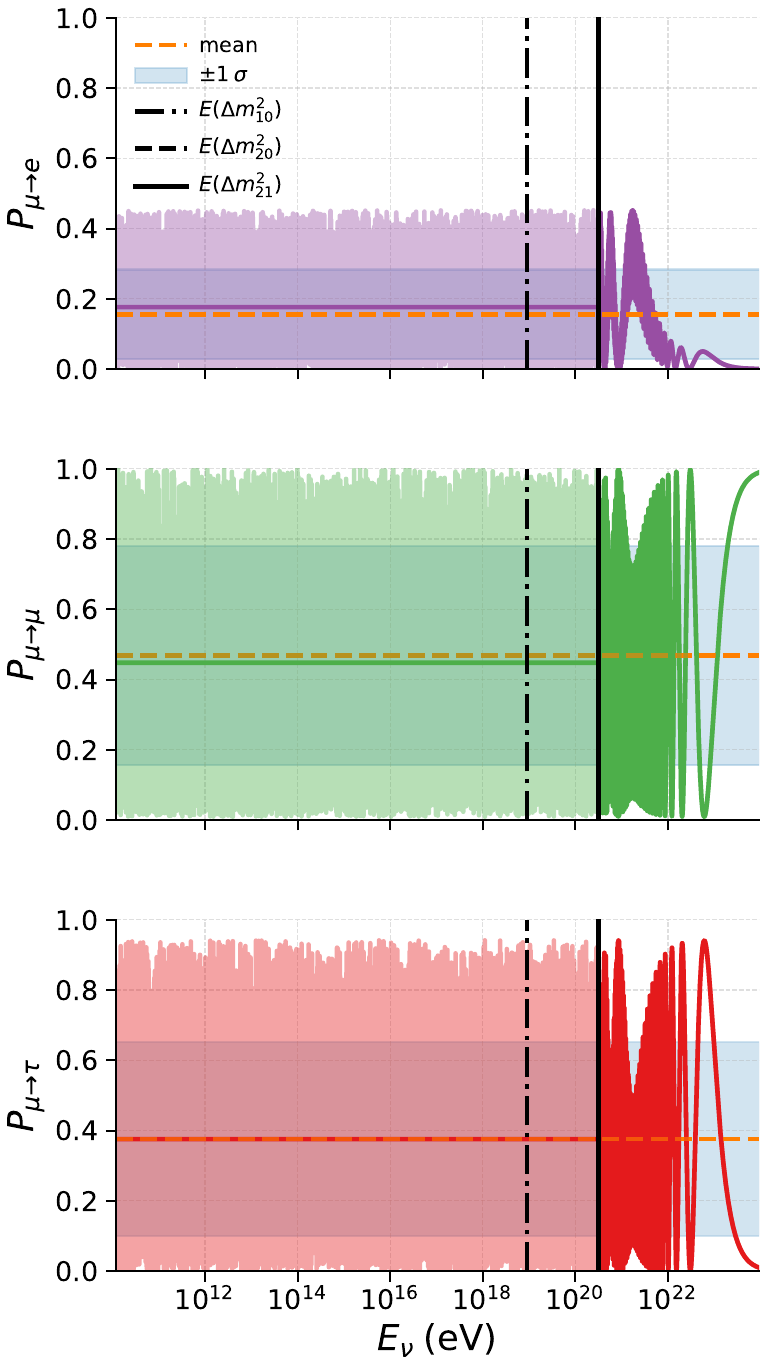}
    \caption{Flavour-transition probabilities for muonic neutrinos of energy $E_{\nu}$ propagating in a 1~parsec region. The phase threshold to shift from the \textit{exact} to the \textit{averaged} regime is set to~300. In the panels the probability mean is also showed together with its dispersion. The vertical lines represents the neutrino energy at the threshold phase for the different mass splitting.}
    \label{fig:oscPlot}
\end{figure}

The~\texttt{$\nu$propa} plug-in mounted on the CRPropa code is composed of three interaction modules, namely \texttt{NeutrinoPhotonInteraction}, \texttt{NeutrinoNeutrino*} and \texttt{NeutrinoAntineutrino*}. A chart about the simulation workflow and structure is provided in Fig.~\ref{fig:chart}. Given the large amount of channels from $\nu\bar{\nu}$ interaction, the latter module is supported by the \texttt{Channels}, employed by the user to (de)activate specific channels, and \texttt{ChannelsBundle}  modules, both used automatically to evaluate the interaction according to the chosen active channels. The channels are intuitively named, as usually done in the CRPropa code, for example ``NeutrinoiAntineutrinojElastic'' or ``NeutrinoAntineutrinoUpResonance''. Due to the dependence of the interaction rates on both neutrino flavours and masses (Eq.~\eqref{eq:IMFPmom}), we describe the propagating neutrinos using the flavour basis, whereas the background ones are defined through their mass eigenstates in the \texttt{NeutrinoBackground} classes, i.e. \texttt{CnuB1}, \texttt{CnuB2} and \texttt{CnuB3}. Thus, to select the interaction rate at each step of the simulation, we assign a mass to the propagating neutrinos and a flavour to the background ones from the state mixing probabilities, properly computed within the \texttt{NeutrinoMixing} class. This latter class has to be initialised in advance in order to build the~PMNS matrix of Eq.~\eqref{eq:PMNSmatrix} through the mixing parameters ($\theta_{12}$, $\theta_{13}$, $\theta_{23}$ and $\delta$) provided by the user. This same mixing class is used as an input by \texttt{NeutrinoOscillation}, capable of performing flavour oscillation in the propagation of highly energetic neutrinos in each simulation step. For a neutrino of energy $E$ travelling a distance $L$, the oscillation is readily computed through the exact oscillation probabilities of Eq.~\eqref{eq:oscilProb}. If the oscillation phases are large, i.e. for $\frac{\Delta m_{ij}^2 L}{4E} \gg 1$, the average values of the sine functions are taken, as shown in Fig.~\ref{fig:oscPlot}. The averaged probability is consistent with the averaged value from the exact formula in Eq.~\eqref{eq:oscilProb}. The transition value is arbitrarily set to~300, coinciding with the correspondent phases for~$\Delta m_{20/21}^{2}$ splittings (see the vertical black lines).   

Once the interaction occurs, the products are generated according to the corresponding differential cross sections in the particle centre of mass. To handle the transformations between the particles centre of mass and the simulation systems the \texttt{RelativisticInteraction} is introduced (see Appendix~\ref{relTreatment} for the details). 

\subsection{Calculation of the differential cross section} \label{sec:dsigma}

As discussed in detail in Section~\ref{sec:nupropa}, and also displayed in Fig.~\ref{fig:chart}, knowledge of the differential cross-section of all interaction channels is a critical input for the propagation code.
During propagation, as outlined in Eq.~\eqref{eq:IMFPmom}, the total cross-section is sufficient to evaluate the occurrence of any specific interaction channel. However, the propagation procedure relies on differential knowledge of these interaction rates such that the kinematic distributions of all secondary particles are accurately described.

\paragraph{Differential cross section implementation.}
The scattering computation is generally performed according to
\beq \label{eq:dsigma}
{\rm d} \sigma = \frac{1}{2 \hat{s}} \overline{\sum} |\mathcal{A}|^2 {\rm d}\Phi_n\,,
\eeq
which includes the flux factor, the summed/averaged squared amplitude, and the final-state phase space. 
The flux factor is trivially $\hat{s} = s_{12} = 2 p_1 \cdot p_2$ for massless incoming particles (photons and/or neutrinos) and the squared amplitudes. 

Following the discussion in Sec~\ref{sec:Interactions}, we consider all tree-level 2-to-2 scattering processes that involve neutrino-neutrino, neutrino-antineutrino, and neutrino-photon interactions.
We further include the $\nu +\gamma \to \ell + W$ channel, as well as the pair production of $W$ or $Z$ bosons in the $\nu + \bar \nu$ annihilation channel.
The calculation of these latter channels requires an on-shell treatment/approximation of $W$ and $Z$ boson production. However, it has the benefit that they can be treated as a simple 2-to-2 process (avoiding numerous complications arising in an off-shell computation).
Before providing channel specific details of the calculation, we recall the general form of the two-body phase space factor as evaluated in the rest frame of two outgoing particles $p_i$ and $p_j$ according to:
\beq \label{eq:PS}
{\rm d}\Phi_2(m_{ij}^2;p_i,p_j) = \frac{\rm d \cos \theta d \phi}{2(4\pi)^2} \, \lambda^{1/2}\!\left(1,\frac{m_i^2}{m_{ij}^2},\frac{m_j^2}{m_{ij}^2}\right)\,,
\eeq
with $\lambda$ denoting the Kallen function, and the variables $\theta$ and $\phi$ denoting polar and azimuthal angles respectively, and  $m_{i(j)}$ denoting the mass of outgoing particle with momentum $p_{i(j)}$.

\paragraph{Squared amplitude evaluation.}
Fig.~\ref{fig:CS_neutrino} displays the total cross section as a function of centre of mass energy for the considered channels.

The process dependent ingredient for the calculation in each of these channels is the corresponding squared amplitude. It is thus necessary to obtain expressions for the squared amplitudes for each of the contributing channels. To achieve this, we have analytically calculated the squared amplitudes making use of the \texttt{FormCalc} package~\cite{Hahn:1998yk,Hahn:2016ebn}, in which the results have been obtained for general fermion masses, and keeping track of complex terms appearing in the result.
The obtained expressions have also been directly compared to numerical results obtained with the package \texttt{Recola2}~\cite{Denner:2017wsf}, finding agreement to machine precision.
By making use of crossing-symmetry, it is possible to obtain all required results from the following subset of 8 channels:
\begin{enumerate}
    \item $\nu_\alpha + \bar \nu_\beta \to \nu_\alpha + \bar \nu_\beta$, for the cases $\alpha = \beta$ and $\alpha \neq \beta$.
    \item $\nu_\alpha + \bar \nu_\beta \to \ell_\alpha + \bar \ell_\beta$, for the cases $\alpha = \beta$ and $\alpha \neq \beta$.
    \item $\nu_\alpha + \bar \nu_\alpha \to f + \bar f$, for all $f \neq \ell_{\alpha}, \nu_{\alpha}$.
    \item $\nu_\alpha + \bar \nu_\alpha \to V + V$, for the cases $V = W, Z$
    \item $\nu + \gamma \to \ell + W$.
\end{enumerate}

The resulting squared amplitudes for these 8 channels are implemented as part of the \texttt{$\nu$propa} plug-in.
The results for all other required channels, such as those for neutrino-neutrino interactions, can be obtained from these or by direct calculation.

\paragraph{A worked example, and a comment on total cross-section.}
We note that analytic formulae for the various channels considered in this work have been previously provided in Ref.~\cite{roulet1993ultrahigh}.
These results have provided an important cross-check of the squared amplitudes which have been required for the differential calculation.
To perform this comparison we have numerically and analytically integrated our squared amplitudes according to Eq.~\eqref{eq:dsigma}.
We give an example of how this is achieved for a single channel, namely $\nu_\alpha + \nu_\alpha \to \nu_\alpha + \nu_\alpha$. 
The leading order (LO) squared amplitude, expanded in the electromagnetic coupling, for this channel is:
\begin{align} \label{eq:nunu}
\sum |\mathcal{A}_{\nu_\alpha \nu_\alpha\to \nu_\alpha \nu_\alpha}|^2 &= 
64 \alpha_{G_F}^2 \pi^2 g_{L,\nu}^2 g_{L,\nu}^{*,2} f(\mu_z,s_{12},s_{13}) f(\mu_z^*,s_{12},s_{13}) \,,\\
	f(\mu_z,s_{12},s_{13})  & = 
\frac{s_{12} \left( 2 \mu_z + s_{12}  \right) }{  \left( \mu_z + s_{12} - s_{13}  \right)  \left( \mu_z +  s_{13}  \right)  }\,,\qquad \mu_Z = m_Z^2 - i \Gamma_Z m_Z\,.
\end{align}
Note that here we define the strength of neutrino coupling via:
\begin{align}
g_{L,\nu} &=  \frac{1}{2 c_W s_W}\,, \qquad 
\alpha_{G_F} = \frac{\sqrt{2}}{\pi} G_{F} | \mu_W^2 s_W^2|\,,
\qquad
s_W^2 = 1 - \frac{\mu_W}{\mu_Z}\,.
\end{align}
The variable $c_W, s_W$ are the (co)sine of the weak mixing angle, $\alpha_{G_F}$ is the electromagnetic coupling defined in the $G_F$ scheme, and $G_F$ is the Fermi constant.
Note that for small values of the ratio $s_{12} / \mu_Z$ this interaction can be effectively described by a contact interaction, and the angular dependence of the interaction is highly suppressed.

Making use of Eq.~\eqref{eq:dsigma}, and ignoring width effects, the total cross-section is:
\begin{align} \nonumber
\sigma_{\nu_1 \nu_1\to \nu_1 \nu_1}(s_{12}) &= \frac{G_F^2}{2\pi} m_Z^2 
	\left(  \frac{s_{12}}{ s_{12} + m_Z^2} + \frac{ 2 m_Z^2 }{ 2 m_Z^2 + s_{12} } \ln\left[ 1 + \frac{ s_{12} }{ m_Z^2} \right] \right) \\
	&\approx \frac{G_F^2}{2\pi} \left(2 s_{12} - 2 \frac{s_{12}^2}{m_Z^2} \right)\,.
\end{align}
In the second line we provide a stable approximation of the result valid for $s_{12} \gg m_Z^2$. In general, we find that such approximations are important for maintaining numerically stable results in the limit $s_{12} \gg m_Z^2$. The results are in agreement with with results of Ref.~\cite{roulet1993ultrahigh}, except for the $\nu \bar \nu \to Z Z$ channel. The resultant expression for this channel are included in the Appendix~\ref{app:tridentNuPhoton}.

\paragraph{A comment on trident production and radiative corrections.}
As part of the propagation code, we have considered only tree-level 2-to-2 scattering processes and have treated the $W$ and $Z$ boson as on-shell for certain production channels.

For example, when describing $\nu + \gamma$ interactions we have considered the 2-to-2 process $\nu + \gamma \to \ell + W$, where the decay of the $W$ boson is treated in a subsequent step in the Narrow Width Approximation (NWA).
This is an approximation for the 2-to-3 off-shell scattering process $\nu + \gamma \to \ell + f + \bar f^{\prime}$, in which $f \bar f^{\prime}$ receives a resonant enhancement for $m_{f \bar f^{\prime}} \sim m_W$.
This approximation simplifies the calculation and allows for a faster numerical evaluation of the (simpler) squared amplitude expression when differential information is required.
In particular, we do not need to consider the full three-body phase-space when describing secondaries during the propagation.
As we treat the decay of the $W$ boson with Pythia, we can also make use of its hadronisation model when considering hadronic decays (see Sec.~\ref{sec:secondaries} below).

In order to justify our approach, we perform the complete 2-to-3 calculation and compare this to the 2-to-2 NWA approximation which we have chosen to adopt for the plug-in. 
Further details on this are given in the Appendix~\ref{app:tridentNuPhoton}. 

\begin{figure}
    \centering
    \includegraphics[width=0.66\linewidth]{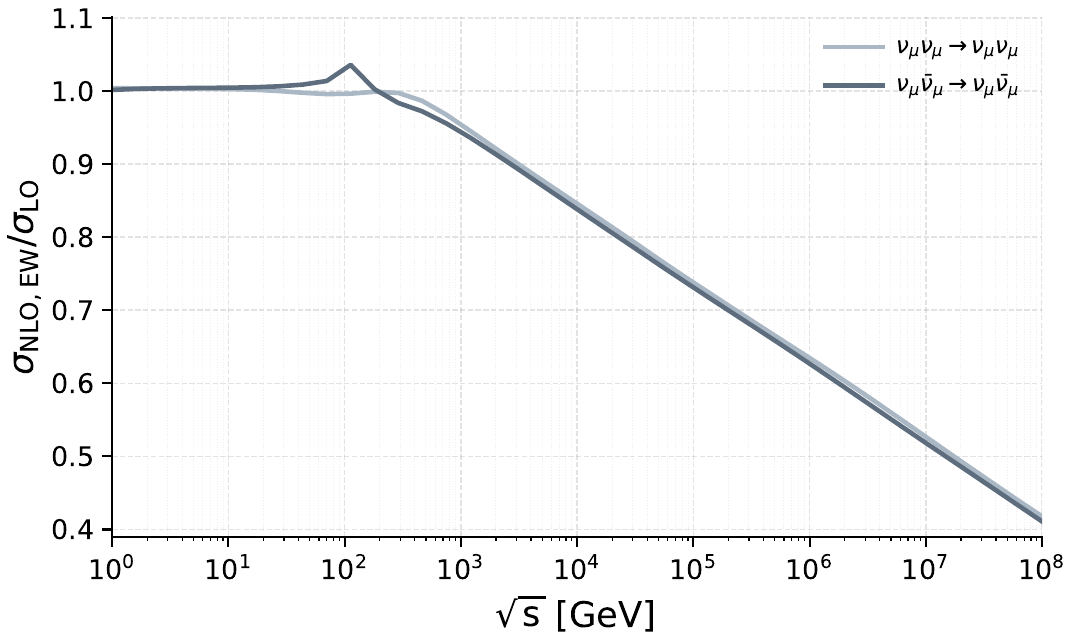}
    \caption{Ratio between the NLO and LO production cross section for the interaction channels $\nu_\mu + \nu_\mu \to \nu_\mu + \nu_\mu$ and $\nu_\mu + \bar \nu_\mu \to \nu_\mu + \bar \nu_\mu$.}
    \label{fig:EWcorr}
\end{figure}

Another approximation adopted in this work is to consider only LO corrections to the various interaction channels, ignoring the role of higher-order Electroweak corrections. Naively, given the smallness of coupling $\alpha$, one expects Electroweak corrections to be of the order of a percent, which goes far beyond the level of precision required for the current analysis. However, as we are considering the propagation of ultra-high-energy neutrinos, it is interesting to investigate the behaviour of these corrections that involve these extreme kinematics.
For example, it is well known that the presence of Sudakov logarithms can lead to large enhancements of the Electroweak corrections~\cite{Sudakov:1954sw}. Although we have not currently included such corrections in the simulations presented here, we thought it was interesting to investigate their potential role for both inclusive and differential cross-section predictions. To achieve this, we numerically compute the next-to-leading (NLO) Electroweak corrections to the process $\nu_\mu + \nu_\mu \to \nu_\mu + \nu_\mu$ (which is also the example considered in Eq.\eqref{eq:nunu}) as well as the resonant process $\nu_\mu + \bar \nu_\mu \to \nu_\mu + \bar \nu_\mu$.
These predictions are obtained with the aid of \texttt{Recola2}~\cite{Denner:2017wsf} to obtain the UV-renormalised virtual corrections and are presented in Fig.~\ref{fig:EWcorr} for the total cross-section.

These results indicate that the NLO EW corrections become significant at extremely large centre of mass energies (reaching $-50\%$).
We note that the numerical evaluation of the NLO virtual corrections is rather time consuming, and therefore directly evaluating the squared amplitude when sampling the differential cross-section for the various interaction channels does not seem tractable for the full propagation framework. Possible future strategies could be to pre-tabulate the NLO predictions for the inclusive and/or differential cross section.
Alternatively, as the dominant numerical contribution of these correction arises from large Sudakov logarithms, it may also be feasible to include analytic approximations for the various interaction channels (at both the differential and inclusive levels)---see, for example, Ref.~\cite{Denner:2000jv,Denner:2001gw,Pagani:2021vyk}. The inclusion of these effects as part of the plug-in is left for future work.

\subsection{Secondaries of interaction} \label{sec:secondaries}
We track the particles produced from the neutrino interactions with the various backgrounds, treating in turn their propagation. The decays of particles like W bosons, muons and tauons are performed in the PYTHIA~8.317 code~\cite{bierlich2022comprehensive}. The plug-in employed is the one implemented in Ref.~\citep{di2025gamma}. Such decays are performed through the \href{https://github.com/GDMarco/CRPYTHIAxDecays}{\texttt{CRPYTHIAxDecays}} plug-in~\cite{di2025gamma}. PYTHIA is also used to perform the hadronization of the quarks produced during the W boson decay, as well as in the resonant production of quark pairs. The outgoing particles are automatically re-injected in the simulation framework and treated as usual CRPropa's \texttt{Candidate}.

As previously shown, e.g.~Fig.~\ref{fig:IMFPnunubarTouubar}, purely hadronic final states are among the possible outcomes from the $\nu\bar{\nu}$ Z-boson resonance interaction. To treat these $q\bar{q}$ final states, the \texttt{$\nu$propa} plug-in is provided by a dedicated \texttt{Hadronisations} module. It generates the outgoing hadrons in a dedicated~PYTHIA class, to then release the stable secondaries in the propagation framework. It also ensures the correctness of the secondary products, possibly discarding the event if the process does not succeed after a few attempts.

\section{Propagation effects on the highest energy neutrinos}\label{sec:exSim}
The relevance of this novel code lies in predicting the actual propagation imprints on the highest neutrino spectra expected from the scenarios introduced in Sec.~\ref{sec:nuProd}. As examples, we simulate simple one-dimensional propagation to emulate the journey of neutrinos through the extragalactic space, choosing step sizes between~$1$ and~$50 \; \text{Mpc}$. The oscillation of the propagating neutrinos is also taken into account: it is initialised with the normal ordering parameters from Ref.~\cite{esteban2024nufit}.  

\begin{figure*}[t]
    \centering

    \subfloat[All-flavour observed neutrino energy spectra. \label{fig:SimuubarNu}]
    {\includegraphics[width=0.48\textwidth]{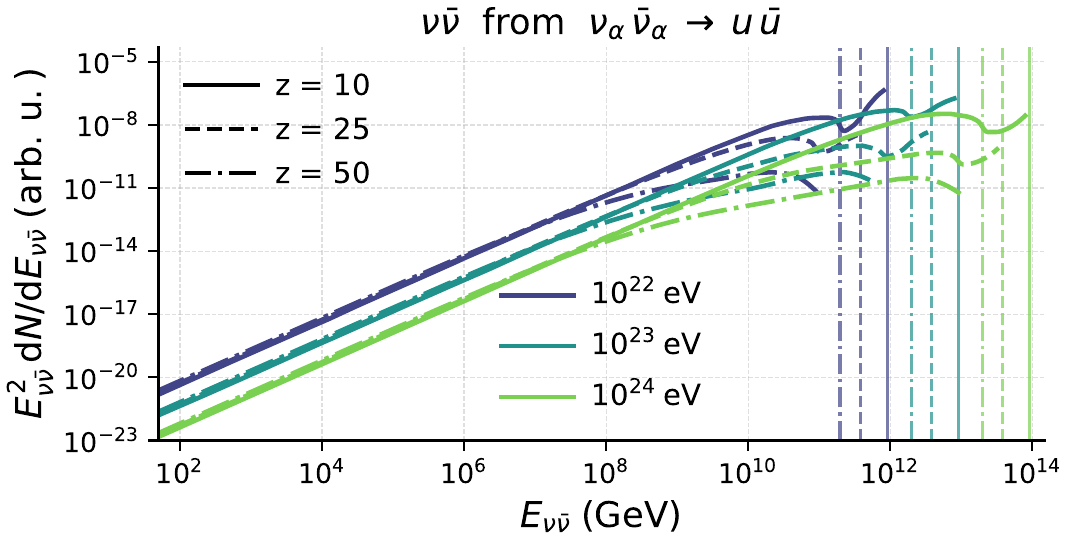}}
    \hfill
    \subfloat[Gamma-ray observed energy spectra. \label{fig:SimuubarGamma}]
    {\includegraphics[width=0.48\textwidth]{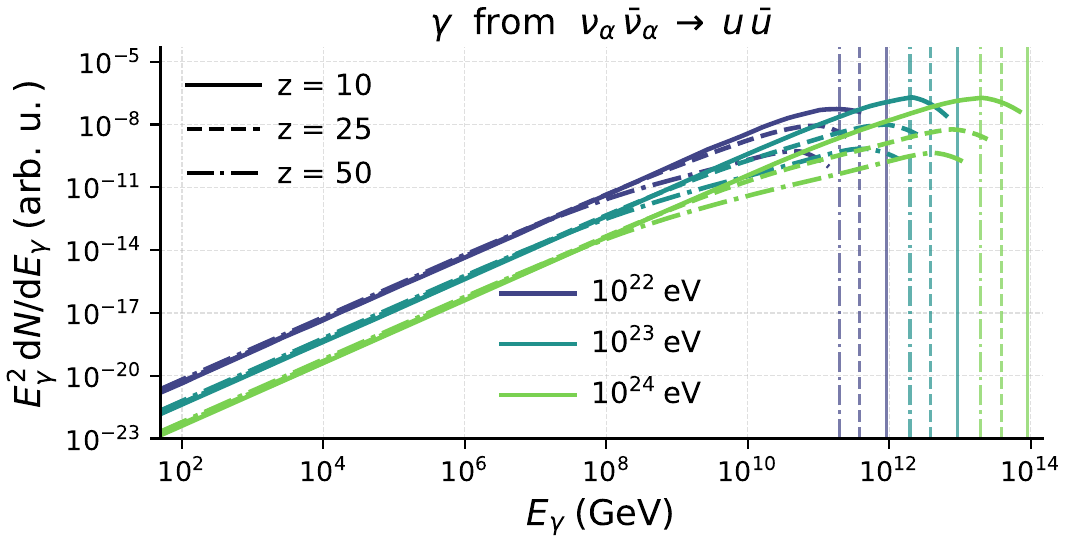}}

    \vspace{0.5cm}

    \subfloat[Electron-positron observed energy spectra. \label{fig:SimuubarEl}]
    {\includegraphics[width=0.48\textwidth]{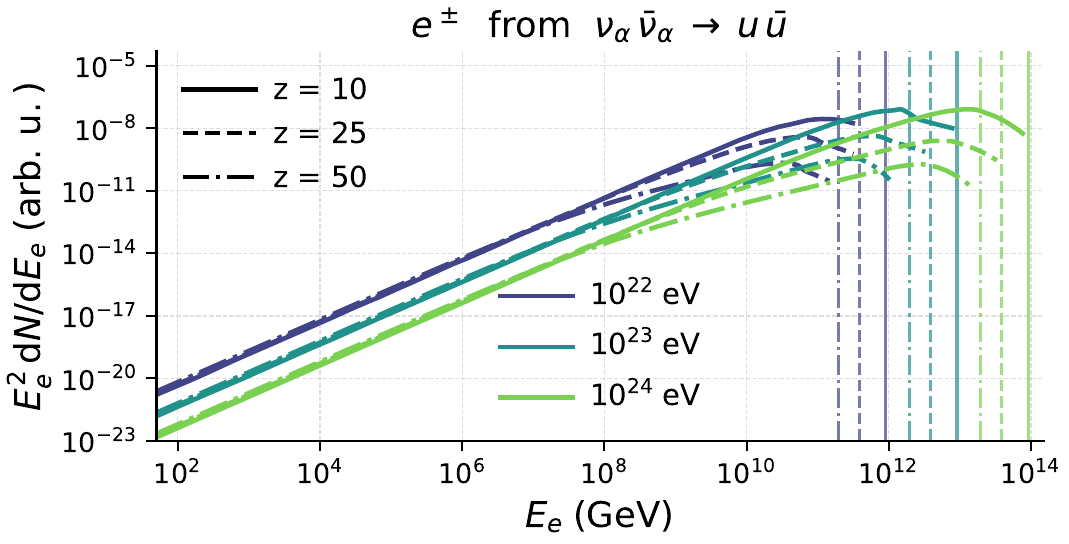}}
    \hfill
    \subfloat[Proton-antiproton observed energy spectra. \label{fig:SimuubarPro}]
    {\includegraphics[width=0.48\textwidth]{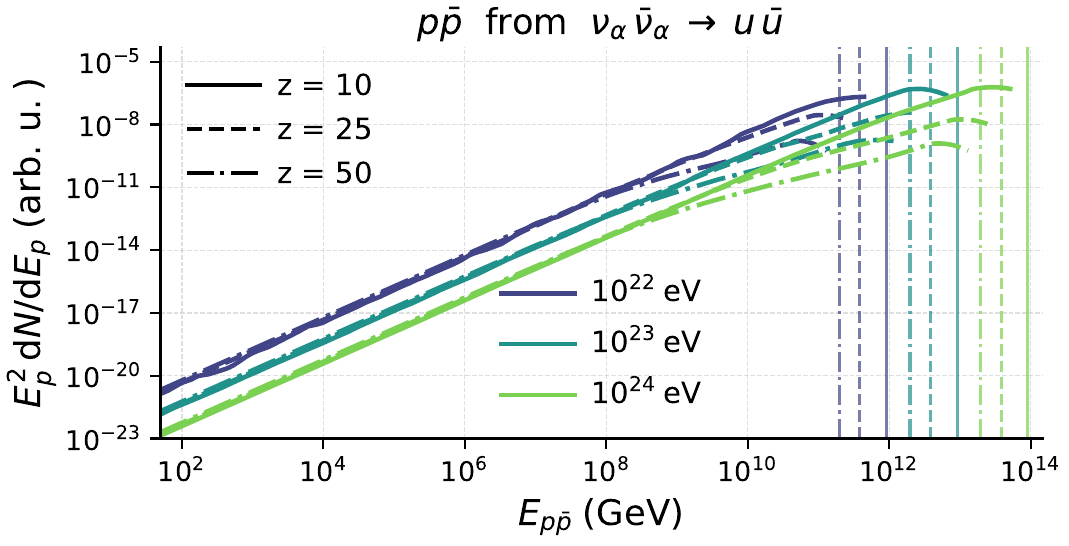}}

    \caption{Energy spectra of the stable messengers produced by the propagation of extremely energetic neutrinos through the C$\nu$B, activating only one interaction channel. The colour legend indicates the energy of the monochromatic sources, while linestyle refers to its redshift, as given in the legend. The dotted vertical lines are the yet redshifted source energies. Each energy spectrum is normalised to its own source energies.}
    \label{fig:spectrauubar}
\end{figure*}

Firstly, we simulate monochromatic neutrino emission pulses occurring at three high redshifts ($z=10,25, 50$). Such emission resembles the decay of a generic heavy relic particle $\chi$ into neutrinos, namely $\chi\to \nu\bar{\nu}$. In these first examples, the only channel in play is the $u\bar{u}$ Z-boson resonance on the C$\nu$B, whose interaction rates are shown in Fig.~\ref{fig:IMFPnunubarTouubar}. The spectra of the outcoming particles, i.e. neutrinos, gamma rays, electrons and protons are in Fig.~\ref{fig:spectrauubar}. Although not included in this example, the resulting electrons and gamma rays might, in turn, interact with the background radiations, initiating electromagnetic cascades in the extragalactic environment~\citep{di2025gamma}. At high redshifts the~CMB will definitely affect their propagation degrading their energies, especially considering the higher order electromagnetic processes as triplet and double pair productions. In the highest energy ranges, also further neutrinos can be \textit{leptonically} produced. At lower redshifts, $z\lesssim4-6$, the~EBL and~CRB would also play an important role, despite the limited knowledge of their densities for $z\gtrsim 1$. Also the protons with energy $\sim 10^{18} \; \text{eV}$ might produce further lower energy fluxes of neutrino and gamma rays by interacting with the background media and photons. The treatment of the cosmological propagation of energetic charged cosmic rays would require assumptions on the pervasive magnetic fields~\citep{hackstein2016propagation, alves2017implications}, in particular the intergalactic one~\citep{durrer2013cosmological}. 

The all-flavour neutrino spectra of Fig.~\ref{fig:SimuubarNu} show the absence of the prompt emission peak for the farthest sources. All the primary neutrinos interact with the C$\nu$B generating flatter and broader observed peaks than the ones from the closer sources. Far from the peak, the spectral slope of the redshift~$z=10$ and~$z=25$ cases start coinciding slightly above $10^{9}\;\text{GeV}$. Below $\sim 10^{8}\;\text{GeV}$ also the spectra from $z=50$ sources align. 

\begin{figure*}[t]
    \centering

    \subfloat[All-flavour observed neutrino energy spectra from just activating $\nu\nu$ (blue line) or $\nu\gamma$ (orange line) interactions. To notice that the solid orange line is not present since no secondary neutrinos are produced in the corresponding case.  \label{fig:SimnunuphNu}]
    {\includegraphics[width=0.48\textwidth]{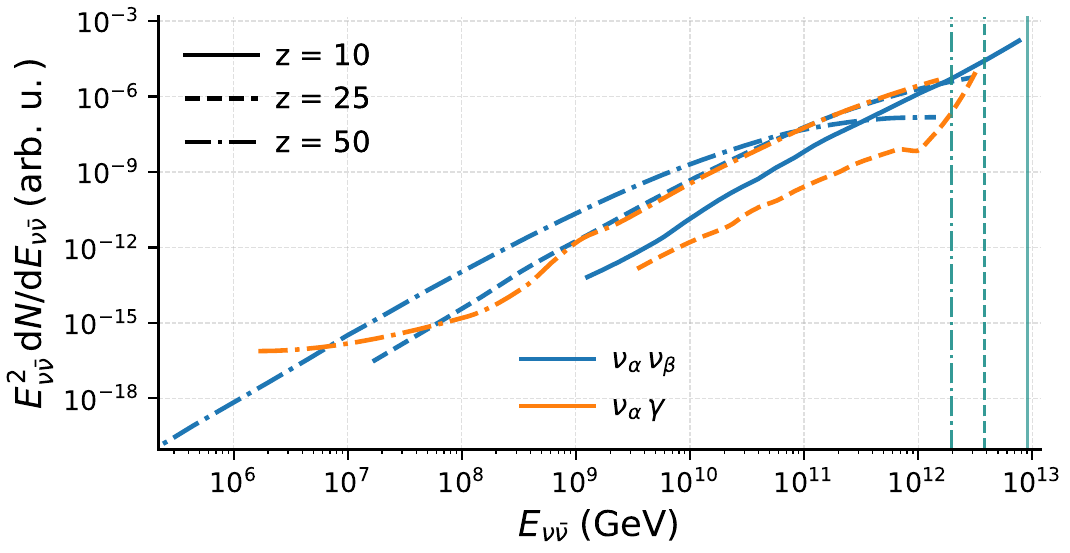}}
    \hfill
    \subfloat[Gamma-ray and electron observed energy spectra from the leptons produced during the propagation through the~CMB photons. \label{fig:SimnuphGammaEl}]
    {\includegraphics[width=0.48\textwidth]{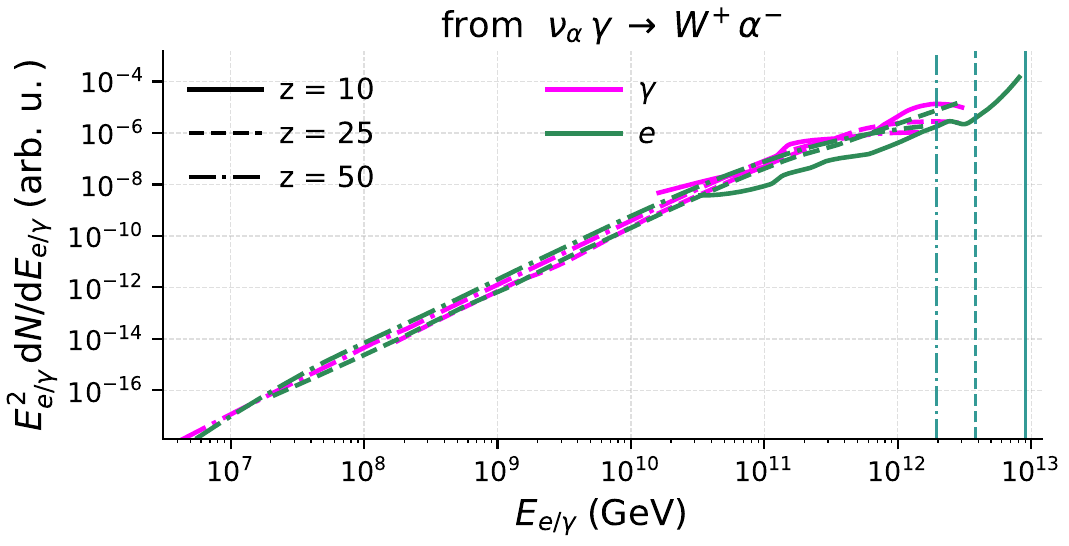}}

    \caption{Energy spectra emerging from monochromatic~$10^{23} \; \text{eV}$ sources. The linestyle denotes the source redshift. The vertical lines are the source energies after the redshift correction, i.e.~$10^{23} /(1 + z)\; \text{eV}$. The normalisation is the same as in Fig.~\ref{fig:spectrauubar}.}
    \label{fig:spectranunuph}
\end{figure*}

We also show the energy spectra at Earth by activating just neutrino--neutrino elastic interactions and neutrino--photon for the uniformly-flavoured intermediate source energy in Fig.~\ref{fig:SimnunuphNu}. Given the energy threshold chosen for these simulations, we do not collect the products from the on-shell~W boson production, but just the ones arising from the companion lepton and/or its decay, as shown in Fig.~\ref{fig:SimnuphGammaEl}.

\begin{figure}
    \centering
    \includegraphics[width=0.66\linewidth]{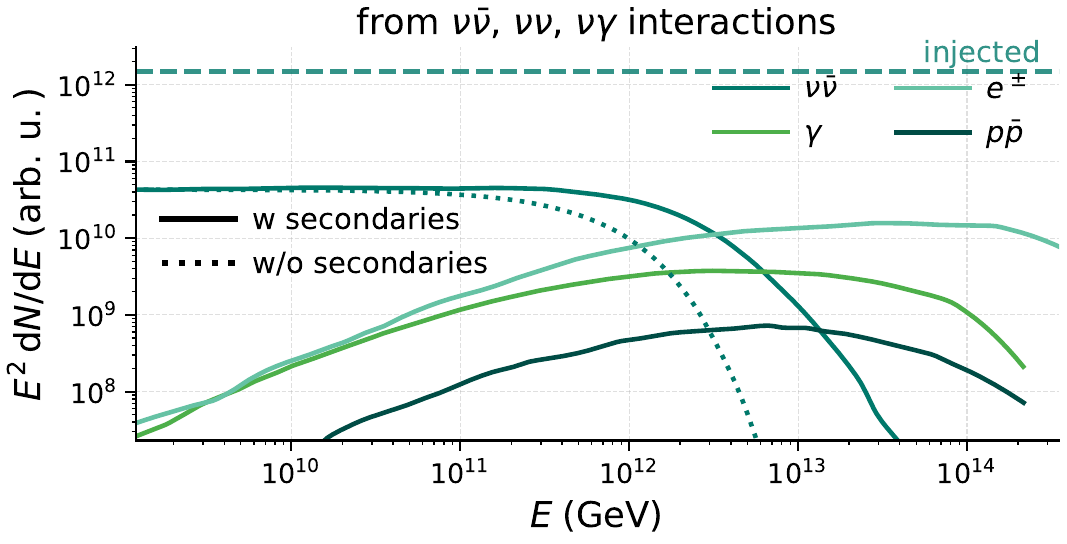}
    \caption{Observed energy spectra from a equal-flavoured neutrino source placed at redshift~10, which emits as a power-law with spectral index~$-2$ in the energy range of the figure (labelled dashed gray line). All the implemented neutrino-(anti)neutrino and neutrino-photon interactions are activated for the three C$\nu$B mass states and the~CMB. The colours of the lines indicate the observed particle, while the linestyle distinguishes between the simulations in which secondaries are traced (solid) or not (dotted).}
    \label{fig:spectraAll}
\end{figure}

We also show in Fig.~\ref{fig:spectraAll} the deformations in the spectra caused by the neutrino--(anti)neutrino and neutrino--photon interactions with the~C$\nu$B and the~CMB. The source is located at~$z=10$ and emits neutrinos through a power-law spectral distribution. The profile of the observed energy spectrum without activating the generation of secondaries (dotted dark teal line) exhibits a smooth decreasing starting from $E_{\nu} \gtrsim 10^{20} \; \text{eV}$. The spectra lowers by three orders of magnitude at~$\sim 6 \times 10^{21} \; \text{eV}$, suppressing the high-energy tail of the emitted neutrinos. The action of the cosmic backgrounds combined with the adiabatic energy losses degrade most of the total injected energy: the ratio between the total observed and injected energy is~$\sim 10^{-4}$. On the other hand, the activation of the secondaries of interaction (solid line) extends the energy spectra towards higher energies, shifting the observed neutrinos by a factor~10 in energy. In this case the observed \textit{versus} injected total neutrino energy is almost the~$5 \; \%$.  The secondary particles overwhelm the primaries reaching the observer both in number and energy. Indeed, more than~210 secondary particles are observed for each injected neutrino. Most of the observed energy comes from, in order of importance, hadronisation processes~($\sim 40\; \%$), direct products of~$\nu\bar{\nu}$ interactions~($\sim 28\; \%$), decays of tauons and muons~($\sim 30\; \%$) and~$\nu\nu$ scatterings~($\sim 2\; \%$). The contributions to the total observed energy from the surviving primary neutrino, products of $\nu\gamma$ interactions and neutron decays are marginal. Among the stable secondary species, the electron--positron component dominates at the highest energies, since it receives contributions not only from hadronisation but also from several leptonic channels.

\begin{figure}
    \centering
    \includegraphics[width=0.66\linewidth]{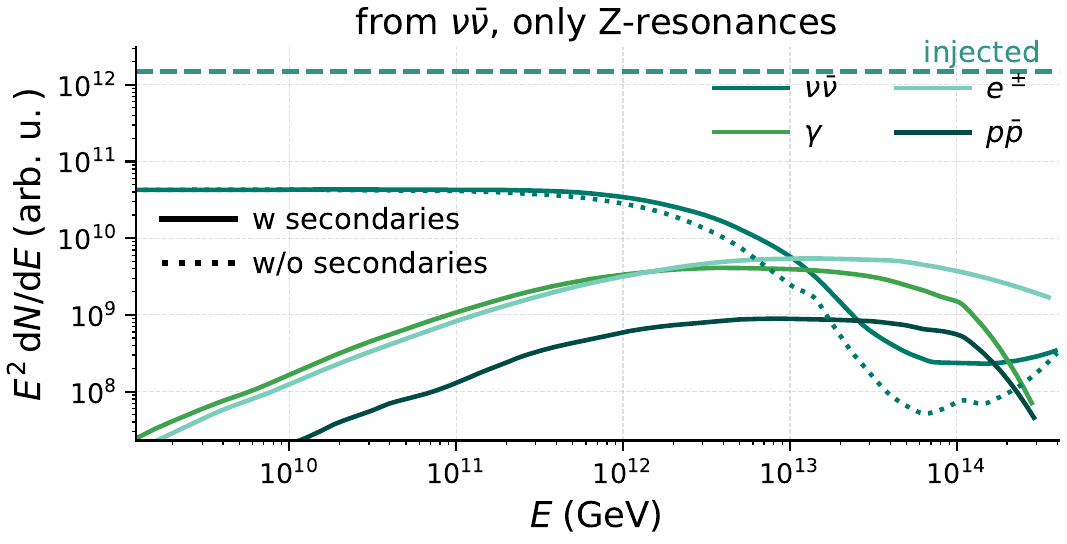}
    \caption{The same energy spectra as Fig.~\ref{fig:spectraAll}, this time activating only the Z-resonant neutrino-antineutrino channels. }
    \label{fig:spectraZres}
\end{figure}

We also simulate the same source as before activating only the neutrino-antineutrino resonant processes. The energy spectra are depicted in Fig.~\ref{fig:spectraZres}. Neglecting the secondaries, the energy spectra start lowering drastically~$\gtrsim 10^{21} \; \text{eV}$, presenting a first feature at~$\sim 2\times10^{22} \; \text{eV}$. The two minima in the spectral profile are situated at~$E_{\nu}\sim 8 \times 10^{22}$ and~$\sim 2\times 10^{23} \; \text{eV}$. These signatures, different in shapes and locations, are due to the peaks in the interactions rates (see, e.g., Fig.~\ref{fig:IMFPnunubarTouubar}) that depend on the background neutrino masses. Despite the differences in background mass and prompt spectrum, they resemble the same Z-boson resonance features found in previous computations (e.g. Fig.~2 of Ref.~\citep{das2024probing}). In the only Z-boson resonances scenario, the total observed energy is a factor~$10^3$ lower than the injected one. By enabling the production of the final states and consequent decays or hadronisations, the neutrino energy spectra remain higher than the previous case by factors of \textit{a few}. The spectral features spotted before are no longer visible. The observed neutrinos carries approximately~$3\; \%$ of the total prompt energy. Most of the observed energy (around the~$79 \; \%$) arises from hadronisation processes; other stable secondaries from $\nu\bar{\nu}$ and from heavy lepton decays carry, respectively,~$\sim 11 \; \%$ and~$\sim 9\; \%$ of the total energy. The energy fraction from the decaying neutrons in the cascade is negligible, while primary neutrinos that are just \textit{redshifted} contributes up to~$1 \; \%$ to the total energy budget. Taking into account the full cascading effect, the observed all-flavour neutrino energy spectrum  stops dropping and reaches a \textit{plateau} at~$\sim 7\times 10^{22} \; \text{eV}$. The comprehensive cascading effects result in a much higher regeneration of neutrinos than previous calculations (the same Fig.~2 of Ref.~\citep{das2024probing} only consider~$\nu\nu$ final states). Also in the case of the Z-boson resonance scenario, the stable leptons detected dominate over the spectra of gamma rays and protons and anti-protons at $E\gtrsim 10^{21} \; \text{eV}$. The gamma-ray and (anti)proton spectra show a suppression at~$\sim 10^{23} \; \text{eV}$.

In general, what it is noticed from the two examples above is that the cascading neutrino energies are strongly degraded in the highest energy regimes, depending on the interactions considered and on the characteristics of the~C$\nu$B backgrounds. However, the comprehensive treatment of the cascades mainly triggered by neutrino-(anti)neutrino interactions makes the observed spectra extending towards higher energies. 


\section{Conclusion \& prospects}\label{sec:conclProsp}

This work aims to present an original simulation framework, the \texttt{$\nu$propa} extension for the~CRPropa code, in which the cosmological propagation of extremely energetic neutrinos is described with unprecedent detail. Since \texttt{$\nu$propa} is mounted and embedded in the CRPropa code, it allows comprehensive characterisation of the secondaries of interactions and, in turn, their propagation through the extragalactic media. The multimessenger studies of the cascades generated by the propagation of cosmological energetic neutrinos, with all the~CRPropa code functionalities -- as, for instance, adiabatic energy losses, magnetic fields, thus~3D characterisation of the charged component of the cascade, and other processes affecting the secondaries -- leads to robust predictions for the neutrino experiments in the highest energy regimes. It is shown, for instance, for the simulated power-law neutrino sources placed at~$z=10$ in Figs.~\ref{fig:spectraAll} and~\ref{fig:spectraZres}. In the former, the activation of the neutrino-(anti)neutrino and neutrino-photon interactions on the~C$\nu$B and~CMB causes a strong absorption for~$E_{\nu} \gtrsim 10^{20} \; \text{eV}$, despite the secondary neutrinos arising from the propagation effects can extend further the energy spectra. In the latter, the Z-boson resonance neutrino-antineutrino processes provoke features in the energy spectra that are dependent on the background mass; the generation of the secondaries of interaction smooths out these traits and increases the observed flux. In general, it is spotted the predominance of the electromagnetic counterpart compared to the baryonic output, although their propagation effects need to be taken into account. The detailed characterisation of the neutrino counterparts is beyond the scopes of this paper and it is left for future detailed studies.     

Throughout the paper, the relevance of the particle backgrounds, i.e.~C$\nu$B and~CMB, is emphasised together with their cosmological evolution and nature. As a matter of fact, extremely energetic neutrinos are not efficiently hindered at low redshifts, thus for lower densities of the backgrounds. However, if such neutrinos are produced farther back in cosmological time, their collisions with the relic backgrounds are more effective and can even reduce almost entirely the prompt flux. However, the energy spectra of the secondary fluxes can peak at lower energies than the prompt component and exhibit an extended low-energy tail. This fact is illustrated in a series if examples of sources located at very high redshift ($\gtrsim 10$) (see Figs.~\ref{fig:spectrauubar} and~\ref{fig:spectranunuph}). Milder effects are expected also for neutrinos produced at more recent cosmic times. Especially when non-relativistic, the C$\nu$B mass affects the profiles of the propagating energetic neutrinos interaction rates. 

Our results showcase the importance of considering all the possible interaction channels and their products: it is what we necessitate to determine reliable flux predictions at Earth in several extreme scenarios. Among them, cosmological fluxes of extremely energetic neutrinos from primordial heavy particles, (super) heavy dark matter decay or cosmic strings coupled to scalar fields. More accurate flux predictions for forthcoming gamma-ray and radio observatories will allow for more robust constraints on such models, as well as better handling of their systematics due to the assumptions on masses and interactions at play~\cite{das2024probing}. Furthermore, investigations on possible distortions of the~C$\nu$B density from the observations of bright neutrino sources, e.g. NGC~1068, can be performed within this new framework~\citep{franklin2024constraints}. 

Our calculations also include a detailed treatment of the secondary particles produced in neutrino-(anti)neutrino and neutrino-photon interactions within a multimessenger framework. These calculations allow us to test the consistency of predicted secondary fluxes with diffuse observations across different messengers, e.g. gamma rays, electrons, and protons.      

The adopted cross section model provides an efficient description of the differential cross-section which underpins these interactions. Importantly, we have verified the accuracy of this model by comparing it to a full treatment of trident production for neutrino-photon interactions and have studied the role of~NLO~EW corrections to neutrino-neutrino interactions at ultra high energies (where large effects, $\approx -50\%$, were observed).
We leave for follow-up work the implementation of these corrections, or approximations of them, in the simulation framework.

Overall, in this article we have shown the importance of accurately modelling the physics of neutrino propagation when studying scenarios that predict neutrino emission beyond the \textit{cosmogenic} regime. Using state-of-the-art calculations and simulation codes, we provide a consistent and exhaustive framework in which current limits on the several extremely energetic models can be revisited and refined, also in further improved models of the interactions, as e.g. shown for the trident $\nu\gamma$, and the oscillation pattern, e.g. accounting for  \textit{decoherence} of states. These predictions will also pave the way for interpretation of forthcoming neutrino radio detections in the highest-energy regimes, where cosmological and particle physics effects meet along their way, altering the overall picture.  

\section*{Acknowledgements}

We thank Kohta Murase, Carlos A. Arg\"uelles, Shigeru Yoshida and the members of the~\hyperlink{https://projects.ift.uam-csic.es/damasco/}{DAMASCO group} at the Institute for Theoretical Physics (IFT) in Madrid for useful discussions. GDM's work is supported by \textit{FPI Severo Ochoa} PRE2022-101820 grant. The work of~GDM and~MASC was supported by the grant PID2024-155874NB-C21. This publication has also been funded within the framework of the R$\&$D$\&$I Project CEX2025-001574-S, funded by MICIU/AEI/10.13039/501100011033. The research presented in this publication falls within the research line 'Origin and Composition of the Universe: Astroparticles and Cosmology (Astro/Cosmo)'. RAB acknowledges the support from the Agence Nationale de la Recherche (ANR), project ANR-23-CPJ1-0103-01. AGS is supported by the CDEIGENT Grant No. CIDEIG/2023/20, by the MICIU/AEI grant PID2024-156285NB-C41, the SO project CEX2023-001292-S, and a 2024 Leonardo Grant from BBVA Foundation.

\section*{Data and code availability}

The data from which the results of this paper are derived are available upon request. The extension for the CRPropa code developed in this work is available at \hyperlink{https://github.com/GDMarco/NuPropa/tree/main}{\texttt{$\nu$propa} github page}. For further details regarding data, analysis and codes used in this work, readers are encouraged to contact the authors. AI assistance was used to implement consistency guards for the hadronisation outputs.

\bibliographystyle{apsrev4-2}
\bibliography{Refs}

\appendix

\section{Further details on cross-section predictions and 2-to-3 scattering}\label{app:tridentNuPhoton} 

In Section~\ref{sec:dsigma} we provide details on the cross-section model used throughout this work and that which is implemented as part of the \texttt{$\nu$propa} plug-in.
This implementation includes analytic results for both the differential and total cross section for all considered 2-to-2 interaction channels.
As part of this we consider the on-shell production of $W$ bosons in neutrino-photon interactions, as well as massive diboson production in neutrino-antineutrino interactions (either $W$ or $Z$ pair production), treating the $W$ and $Z$ bosons in the NWA.

In this Appendix we give further details on two aspects of these calculations.
First, since we do not find agreement with the analytic result provided in~\cite{roulet1993ultrahigh} for the $\nu + \bar \nu \to Z + Z$ cross section, we give our analytic result for this interaction channel.
Secondly, we also provide details on an independent off-shell calculation of 2-to-3 scattering processes in neutrino-photon interactions.
While the results of the latter have not been included as part of the plug-in, they were used to validate the approximation which has been adopted.
Furthermore, we comment on potential strategies to include differential scattering predictions as part of the propagation code that go beyond the NWA.

\paragraph{Massive diboson production.}
As discussed in Section~\ref{sec:dsigma}, we have implemented analytic results for the differential cross section for each channel.
Analytic expressions for the total cross section were previously provided in~\cite{roulet1993ultrahigh}.
As a cross-check of our calculation of the differential cross section, we analytically integrate all of our expressions and compare them to the existing results in the literature.
Adopting the same Electroweak input scheme as in~\cite{roulet1993ultrahigh} we found agreement for all channels except for the $\nu + \bar \nu \to Z + Z$ process. We therefore quote the results we found for this channel:
\begin{align} \nonumber
\sigma_{\nu \bar \nu \to Z Z}(s_{12})  &=\frac{8 \alpha_{G_F}^2 \pi g_{L,\nu}^4}{s_{12} (y-2) y}
\left(  \frac{1}{2} (y^2+4) \ln \left[ \frac{a+1}{a-1}  \right] - \sqrt{(y-4)y^3} + 2 \sqrt{(y-4)y} \right) \\
y &= \frac{s_{12}}{m_Z^2} \,,\qquad a = \frac{y-2}{\sqrt{(y-4)y}} \,.
\end{align}
To validate this result we first performed a numerical check of the squared amplitude against that obtained with \texttt{Recola2}, and then performed both a numerical and analytic integration of phase-space.
We obtained consistent results when taking either approach.

\paragraph{Neutrino-photon interactions and off-shell effects.}
As previously discussed, we include neutrino-photon interactions in our calculation by considering on-shell $W$ boson production.
Namely, we consider the 2-to-2 interaction $\nu_{\ell} + \gamma \to \ell + W$, and then later consider the decay of the $W$ boson in the NWA (through the interface to Pythia mentioned in Sec.~\ref{sec:secondaries})
Clearly, this is a more simple and numerically efficient approach than considering the full off-shell 2-to-3 scattering process.
In addition, through the interface to Pythia, it allows for the inclusion of hadronisation effects when considering the secondaries.

To benchmark/compare how well the 2-to-2 NWA approximation holds, we have performed a calculation of the full off-shell 2-to-3 process for all possible channels.
To perform this calculation we have made use of the existing real-emission squared amplitudes which were required for the NLO EW calculation of the Glashow resonance~\cite{Gauld:2019pgt}, which were implemented in an independent code to allow for a numerical calculation of the differential cross-section according to Eq.~\eqref{eq:dsigma}.
The required three-body phase space is implemented using a recursion of the two-body phase space appearing in Eq.~\eqref{eq:PS}. Namely
\beq
{\rm d}\Phi_3(m_{ijk}^2;p_i,p_j,p_k) = {\rm d}\Phi_2(m_{ijk}^2;p_i,p_{jk}) \frac{d m_{jk}^2}{2 \pi} {\rm d}\Phi_2(m_{jk}^2;p_j,p_k)\,.
\eeq
With those ingredients in place, it is straightforward to perform a numerical calculation of the differential or total cross section.
This has been achieved by making use of the {\tt VEGAS} algorithm that is implemented in the {\tt CUBA} library~\cite{Hahn:2004fe,Hahn:2014fua}.

\begin{figure*}[t]
    \centering
    \includegraphics[width=0.68\textwidth]{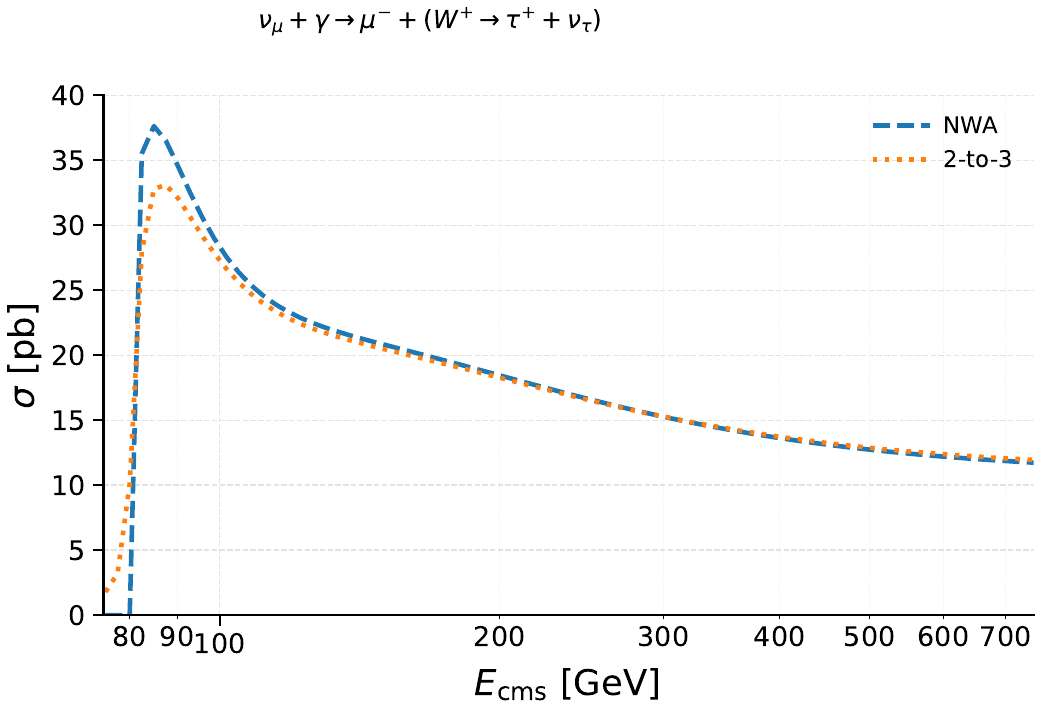}
    \caption{Comparison of the total cross section for the interaction channel $\nu_{\mu} + \gamma \to \mu^- + \nu_{\tau} + \tau^+$, considering the NWA (blue dashed) and the off-shell 2-to-3 calculation (orange dotted). \label{fig:trident}}
\end{figure*}

To compare the two approaches, we have numerically evaluated the total cross section for the channel $\nu_{\mu} + \gamma \to \mu^- + \nu_{\tau} + \tau^+$, and for $\nu_{\mu} + \gamma \to \mu^- + (W^+\to \nu_{\tau} + \tau^+)$ where the $W$ boson is considered in NWA.
A result of these calculations is shown in Fig.~\ref{fig:trident}.
Close to the resonance $E_{cms}\sim m_W$ there are differences in the predicted rate as the off-shell calculation is distributed according to the Breit-Wigner distribution in this region, while the NWA includes a sharp threshold at $E_{cms}=m_W+m_{\mu}$.
For $E_{cms}\gtrsim 100$~GeV the calculations lead to a very similar result.
The off-set of the two calculations in this region is at the percent level and is, in part, related to the choice of the branching faction applied in the NWA calculation (1/9) versus the input value of $\Gamma_W$ which enters the off-shell calculation.

One final point we wish to make is that while the NWA provides a good description of the overall rate, it neglects non-resonant contributions such as interactions of the form $\nu + \gamma \to \nu + f \bar f$.
These contributions are typically numerically small and have not been included in the propagation framework.

\section{Relativistic treatment in the secondaries production} \label{relTreatment}

In CRPropa, at each propagation step, the occurrence of a certain scattering is evaluated through MC tools according to the interaction rates. In generating the secondaries of interaction, the differential cross sections in the centre of mass frame are employed. The latter is computed from the squared centre of mass interaction energy, $s$, and the angle between the interacting particle, called by the subindex 1, and a product, 3, thus $\theta_{13}^{*}$.  

\paragraph{Interactions between neutrinos.}

From the (precomputed) differential rate distribution, $d\lambda / d s $, and the energy of the propagating neutrino, $E_{\nu}$, the $s$ of the interaction is selected. Since the isotropic orientation of the C$\nu$B, the cosine of the interaction angle $\cos\theta$ in the plane where two momenta lie, is chosen randomly between -1 and 1. The centre of mass energy is given by: 
\begin{equation}\label{eq:comEn}
    E^{*} = \Big(m_{\nu}^{2}c^{4} + m_{\text{b}}^{2}c^{4} + 2 E_{\nu}\epsilon-2E_{\nu}\sqrt{\epsilon^{2}-m_{\text{b}}^{2}c^{4}}\cos{\theta} \Big)^{1/2} \,,
\end{equation}
considering $E_{\nu}\sim cp_{\nu}$. From Eq.~\eqref{eq:comEn}, the energy of the background neutrino is readily computed as:
\begin{equation}
    \epsilon = \frac{K_{1}+\sqrt{4K_{2}}}{2E_{\nu}(1-\cos^{2}{\theta})}
\end{equation}
in which the constants are:
\begin{equation}
    K_{1} \equiv s - (m_{\nu}^2 +m_{\text{b}}^2) c^{4} \,,
\end{equation}
\begin{equation}
    K_{2} \equiv K_{1}^{2}/4 + E_{\nu}^{2}m_{\text{b}}^{2}c^{4}\cos^{2}{\theta} \,.
\end{equation}

Thus, we define the centre of mass velocity along the direction of propagation of the highly-energetic neutrino, viz. a generic $z$--axis, as: 

\begin{equation}
\beta^{*}_{z} = \frac{\sqrt{E_{\nu}^{2}-m_{\nu}^{2}} + \sqrt{\epsilon^{2}-m_{\text{b}}^{2}}\cos{\theta}}{E_{\nu} + E_{\text{b}}} \,,
\end{equation}
with the corresponding Lorentz factor:
\begin{equation}
\gamma^{*} = \frac{E_{\nu} + \epsilon}{\sqrt{s}} 
\end{equation}
in which $\sqrt{s} \equiv E^{*}$, the centre of mass energy. The other components of the centre of mass velocity, i.e. ($\beta_{x}^{*},\beta_{y}^{*}$), are negligible since $\propto \epsilon / E_{\nu}\ll1$.

The particles produced by $2 \to 2$ scatterings, with masses $m_{3}$ and $m_{4}$, are produced in the centre of mass with the same momenta, i.e. $p^{*} \equiv |\vec{p}^{*}_{3}| = |\vec{p}^{*}_{4}|$ and angles with respect to the incoming energetic neutrino $\theta_{13}^{*}$ and $\theta_{14}^{*} = \theta_{13}^{*} + \pi$. 
The angle $\theta_{13}^{*}$ is drawn from the function for the differential cross section according to
\begin{equation} 
\frac{d\sigma^{*}} {d \cos{\theta_{13}^{*}}} (s, \theta_{13}^{*}) \,, 
\end{equation}
which is pre-tabulated for efficiency. In this frame, as the z-axis has been aligned with the direction of the highly-energy  neutrino, the azimuthal integration around the z-axis in the starred (centre of mass frame) that appears in the two-body phase-space (see Eq.~\eqref{eq:PS}) is trivial. 

From $s = (E_{3}^{*}+E_{4}^{*})^{2}$ and the relativistic dispersion equations, the momentum in the centre of mass is calculated as:
\begin{equation}
    p^{*} = (s/4 - (m_{3}^{2}+m_{4}^{2})/2 + (m_{3}^{2}-m_{4}^{2})^{2}/4s )^{1/2} \,,
\end{equation}
given that $s\geq s_{\text{thr}}=(m_{3}^{2}+m_{4}^{2})$. Once computed the product energies in the starred frame, the products energies are obtained by boosting back to the simulation frame along the $z$-axis: 
\begin{equation}
    E_{3} = \gamma^{*}(E_{3}^{*} + \beta^{*}_{z}p^{*}\cos\theta_{13}^{*}) \,, 
\end{equation}
\begin{equation}
    E_{4} = \gamma^{*}(E_{4}^{*} + \beta^{*}_{z}p^{*}\cos(\pi +\theta_{13}^{*})) \,.
\end{equation}
The interaction products are emitted along the same direction of the progenitor energetic neutrino.

\paragraph{Neutrino-photon interactions.}
In the case of energetic neutrinos interacting with CMB photons, for instance, $s$ and $\theta$ are selected as in the previous paragraph. In this case the photon energy is worked out as:
\begin{equation}
    \epsilon_{\gamma} = \frac{s - m_{\nu}^{2}}{2E_{\nu}(1-\cos{\theta})} \,.
\end{equation}
The $z$ component of the centre of mass velocity:
\begin{equation}
    \beta^{*}_{z} = \frac{\sqrt{E_{\nu}^{2}-m_{\nu}^{2}} + \epsilon_{\gamma}\cdot\cos{\theta}}{E_{\nu_{\text{UHE}}} + \epsilon_{\gamma}} \,,
\end{equation}
and the corresponding Lorentz factor:
\begin{equation}
    \gamma^{*} = \frac{E_{\gamma} + E_{\nu}}{\sqrt{s}} \,.
\end{equation}
The product energy is evaluated as in the previous paragraph.

\end{document}